\documentclass[%
aps,
pra,%
amsmath,amssymb,showpacs,
reprint,%
floatfix,
twocolumn,
nofootinbib,
longbibliography,
superscriptaddress
]{revtex4-2}

\usepackage[dvipsnames]{xcolor}

\usepackage[]{todonotes}
\usepackage{amsmath,amsfonts,amssymb}
\usepackage{cases} 
\usepackage{xfrac}

\usepackage{graphicx}

\usepackage{hyperref}

\allowdisplaybreaks[3]

\newcommand{\newt}[1]{{\color{black}#1}}

\DeclareMathAlphabet\mathbfcal{OMS}{cmsy}{b}{n}

\usepackage{xr}
\newcommand{\figLettInv}{4}

\begin{document}
	
	\title{Excitation of narrow x-ray transitions in thin-film cavities by focused pulses}
	
	\author{Dominik \surname{Lentrodt}}
	\email[]{dominik.lentrodt@physik.uni-freiburg.de}
	
	\affiliation{Max-Planck-Institut f\"ur Kernphysik, Saupfercheckweg 1, 69117 Heidelberg, Germany}
	\affiliation{Physikalisches Institut, Albert-Ludwigs-Universit\"at Freiburg, Hermann-Herder-Stra{\ss}e 3, D-79104 Freiburg, Germany}
	\affiliation{EUCOR Centre for Quantum Science and Quantum Computing, Albert-Ludwigs-Universit\"at Freiburg, Hermann-Herder-Stra{\ss}e 3, D-79104 Freiburg, Germany}
	
	\author{Christoph H.~\surname{Keitel}}
	
	\author{J\"org \surname{Evers}}
	\email[]{joerg.evers@mpi-hd.mpg.de}
	
	\affiliation{Max-Planck-Institut f\"ur Kernphysik, Saupfercheckweg 1, 69117 Heidelberg, Germany}
	
	
	\begin{abstract}
		A method to compute the excitation of narrow transitions at hard x-ray energies by short focused x-ray pulses is developed. In particular, the effect of thin-film cavities on the pulse propagation is incorporated via a semi-analytical algorithm requiring the numerical evaluation of only one two-dimensional Fourier transform. We investigate various limiting cases to confirm the reliability of the algorithm. As an application, we show how a focused x-ray pulse propagates in cavity structures utilized in previous theoretical studies and experiments with collimated beams.
	\end{abstract}
	
	\maketitle
	
	\section{Introduction}
	In the field of quantum optics, one of the key goals is to gain increasingly detailed control over even individual quantum systems through the use of light fields. For visible and microwave light, many such methods are available by now \cite{Haroche2013,Wineland2013} and various types of excitations in atoms \cite{Chu2002,Ludlow2015}, molecules \cite{Shapiro2011,Koch2019}, nuclear spins \cite{Vandersypen2005} or materials \cite{DeLaTorre2021} can be controlled to a surprising degree.
	
	At higher energies, however, the tools accessible to quantum opticians are much scarcer. In particular in the hard x-ray or $\gamma$-ray domain, it is difficult to create and shape light fields with the necessary intensities, monochromaticity and coherence properties. While recently much progress has been achieved in x-ray source technology \cite{Pellegrini2016,Georgescu2020,Nam2021,Liu2023} and optics \cite{Shvydko2004_book,Salditt2020}, quantum optics with narrow x-ray transitions --- such as M\"ossbauer nuclei \cite{Rohlsberger2021} or inner-shell electronic excitations \cite{Haber2019,Huang2021,Gu2021} --- is still in its infancy.
	
	In the last decade, a relatively new field that has seen a particular surge in experimental and theoretical activity is x-ray cavity quantum electrodynamics (QED). Initiated by the discovery of the collective Lamb shift \cite{Rohlsberger2012} of an ensemble of M\"ossbauer nuclei embedded in a thin-film cavity, a number of interesting quantum optical effects have been realized and predicted, including electromagnetically induced transparency (EIT) \cite{Rohlsberger2012}, spontaneously-generated coherences (SGC) \cite{Heeg2013a}, Fano resonances \cite{Heeg2015a}, slow x-ray light \cite{Heeg2015b,Kong2016}, strong inter-ensemble coupling \cite{Haber2016a,Haber2017} and multi-mode cavity interactions~\cite{Lentrodt2023}. On the theory side, initial phenomenological models \cite{Heeg2013b,Heeg2015c} have been replaced by first principle methods \cite{Lentrodt2020a}, such that a solid understanding of the underlying processes is in principle available and one can even optimize their interplay via the cavity environment \cite{Schenk2011PhD,Diekmann2022,Diekmann2022b,Li2022}. Recently, similar effects have been realized with electronic resonances \cite{Haber2019,Huang2021,ZiRu2022} and the theoretical methodology has been extended to include such systems~\cite{Gu2021,Gu2021a}. Furthermore, x-ray waveguides \cite{Salditt2020,Salditt2008} have been considered as a novel photonic environment to control nuclear transitions~\cite{Chen2022,Andrejic2023_arxiv,Lohse2024_arxiv,Lohse2024a} and x-ray emission from electronic processes \cite{Vassholz2021}.
	
	While this progress establishes narrow x-ray transitions as a new platform for quantum optics \cite{Adams2013} and cavity QED \cite{Rohlsberger2021}, so far, these experiments and also the theoretical descriptions focus on the low excitation sector currently accessible at source facilities \cite{Adams2012,Adams2019} and do not consider focusing of the x-ray pulses. As such, the reported effects were all characterized via linear optics observables \cite{Rohlsberger2005,HeegPhD} and can be captured by linear dispersion theory \cite{Mandel1995,Zhu1990,Andreoli2021} in combination with Parratt's \cite{Parratt1954} or the transfer matrix formalism \cite{Abeles1950,Rohlsberger2005} for the classical wave propagation in the dielectric cavity environment (see \cite{Rohlsberger2021,LentrodtPhD} for recent reviews).
	
	Here, we present a theory which includes the effect of focusing on the pulse propagation and is able to describe the potentially nonlinear excitation dynamics. Specifically, we focus on the transition dynamics during an ultra-short driving pulse introducing an approximation hierarchy which differs from previous Maxwell-Bloch approaches \cite{Junker2012,Liao2012,LiaoPhD}. Instead, we only investigate the excitation dynamics --- particularly considering the parameters regimes relevant for Mössbauer nuclei, but also applicable to similar electronic transitions --- during an ultra-short driving pulse on the fs to ps scale, similarly to approximations which are standard at lower frequencies (see e.g.~\cite{Ames2022}). We note that an alternative approach to nonlinear excitation dynamics has recently been presented in the context of x-ray waveguides in front-coupling geometry \cite{Chen2022}, which utilizes full Maxwell-Bloch equations but ignores the focusing properties of the incident beam.
	
	To this end, we develop a semi-analytical method to include the effects of a focused incident beam in combination with a thin-film x-ray cavity to modify its propagation via interface reflections, absorption and interference \cite{NovotnyHecht2006}. The algorithm makes use of the cavity response being available analytically for a highly collimated input \cite{Parratt1954,Abeles1950,Tomas1995}, which is encoded in existing software packages \cite{Sturhahn2000,pynuss,Lentrodt2022_pyrot,Nexus}. With an appropriate analytical treatment of the incident beam, which we assume to be Gaussian \cite{NovotnyHecht2006}, the resulting problem of a focused beam in the cavity can be reduced to one two-dimensional Fourier transform, which can straightforwardly be performed numerically. The algorithm is therefore highly numerically efficient and accurate, which allows to optimize the cavity geometry for specific purposes, which we exploit in a companion paper~\cite{companion_nonlin2024_arxivLetter} to investigate the feasibility of inverting nuclear ensembles at current x-ray source facilities. Besides nonlinear excitation dynamics, these results open new theoretical options for exploitation of focusing for the investigation of small targets or for depth-selective driving allowing the realization of novel quantum optical effects not realizable with collimated driving.
	
	Sec.~\ref{sec::nqd} outlines how the nuclear dynamics can be solved on different time scales. During an ultra-short driving pulse, optical Bloch equations for a single nucleus are obtained. Sec.~\ref{sec::excDyn_pulseArea} shows how the resulting equations can be solved via the pulse area theorem, introducing a particular Fourier component of the field as the central quantity determining the nuclear excitation directly after the pulse. In Sec.~\ref{sec::exPh_beamDiv}, we show how to treat the propagation of the focused pulse in the x-ray cavity environment. Various benchmarks are presented to confirm the validity of the algorithm. 
	
	\begin{figure}[t]
		\centering
		\includegraphics[trim={2.0cm 0cm 0 1.88cm},clip,width=1.0\columnwidth]{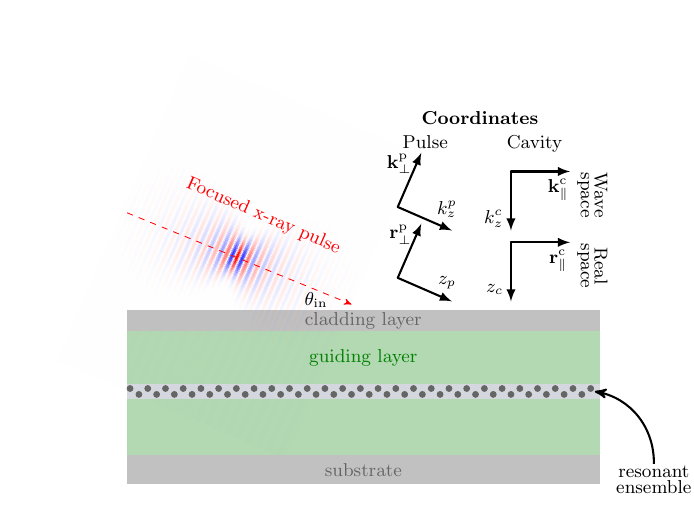}
		\caption{Sketch of the setup, featuring a focused x-ray pulse incident on a thin-film cavity doped with narrow resonances, such as M\"ossbauer nuclei or electronic transitions. The different coordinate systems used in the main text are indicated.}
		\label{fig::examples_illu_setup}
	\end{figure}
    \begin{figure*}[t]
		\centering
		\includegraphics[trim={6.0cm 5.0cm 12.0cm 14cm},clip,width=1.0\textwidth]{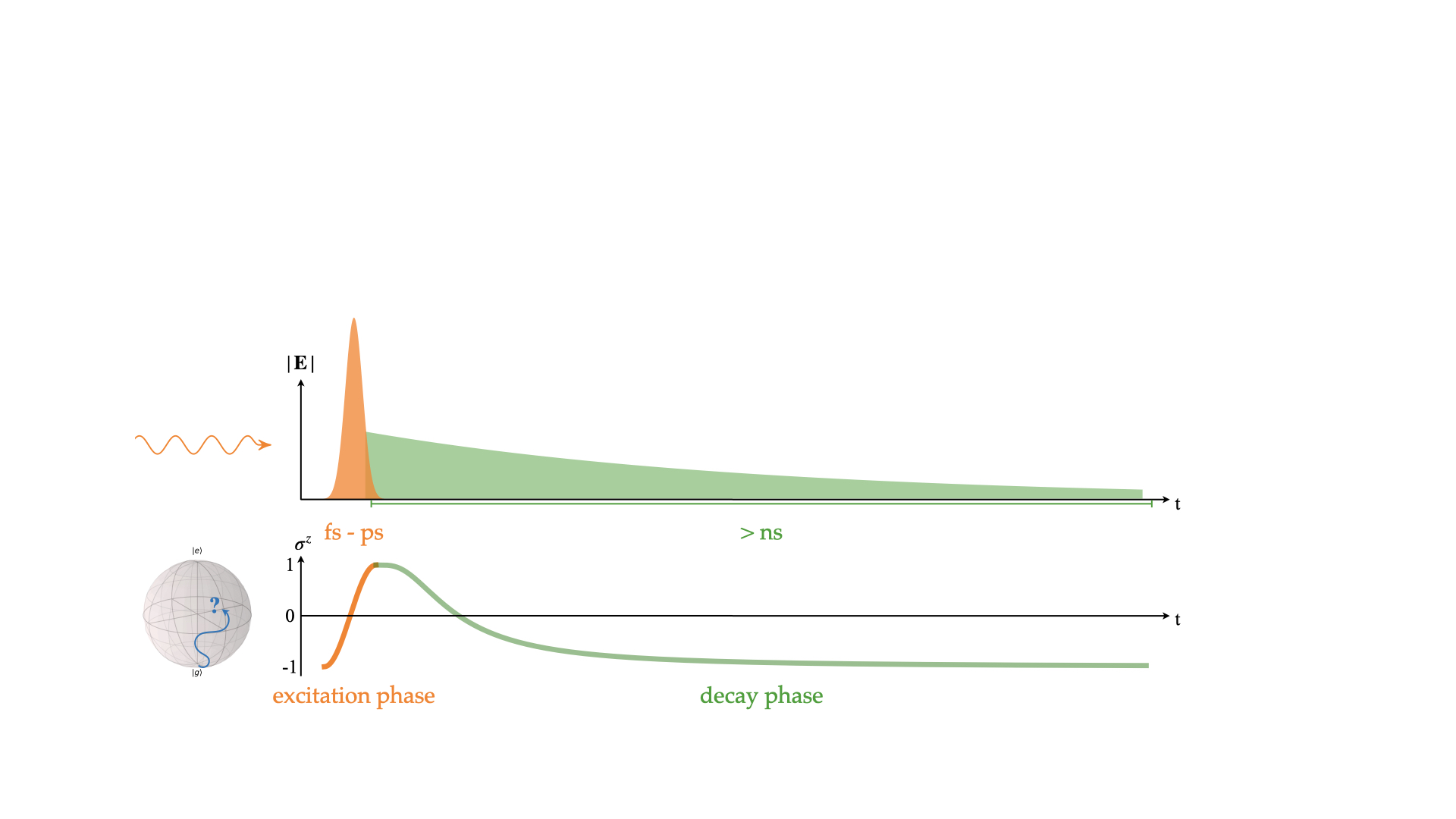}
		\caption{\newt{Illustration of the time scale separation. During the short x-ray pulse on the femto to picosecond scale (top panel: sketch of electric field), the narrow x-ray transition (nuclear or electronic) is strongly driven. We refer to this process as the ``excitation phase'', where the dynamics (bottom panel: sketch of the corresponding $\sigma^z$ of the transition) are solvable since interactions can be neglected. On longer time scales (typically nanoseconds), the transitions then decay and interact with each other, leading to many-body dynamics. The figure shows a sketch (not to scale) of the deexcitation during this ``decay phase''. We note that the time structure of the pulse is not significantly modified by the addition of an x-ray cavity, due to the frequency response of the latter being constant on the scale of the pulse bandwidth.}}
		\label{fig::examples_illu_phases}
	\end{figure*}
	\begin{figure}[t]
		\centering
		\includegraphics[trim={12.0cm 5.9cm 9.0cm 5.5cm},clip,width=1.0\columnwidth]{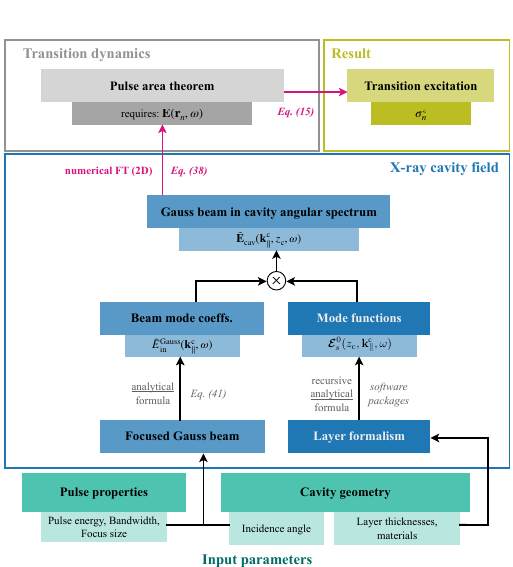}
		\caption{Numerical algorithm developed in this paper, consisting of a single 2D Fourier transform and the area theorem (Eqs.~\eqref{eq::cav_resp_RespFunsExp} and \eqref{eq::pulse_area_solz_phi}, respectively; magenta arrows). The x-ray cavity field is computed via its angular spectrum, which is obtained from the beam mode coefficients (quantifying the input beam in cavity coordinates) and the mode functions (quantifying the response of the cavity). The analytical part of this paper serves to calculate the former for the special gase of a Gaussian beam (see also Appendix~\ref{app:overview} for an extended illustration), the latter can be obtained using existing software packages \cite{Sturhahn2000,pynuss,Nexus}, which implement the so-called layer formalism \cite{Rohlsberger2005}. Note that the numerical part of the algorithm can also be applied to non-Gaussian input profiles, if an appropriate $E_\textrm{in}(\mathbf{k}_\parallel, \omega)$ is provided.}
		\label{fig::examples_illu_algo}
	\end{figure}
	
	\section{Excitation dynamics of the nuclear ensemble transition dynamics}\label{sec::nqd}
	We start by summarizing the theoretical background that our calculation is based on, outlining the model for the nuclear quantum dynamics. We then derive an approximate analytical solution for the excitation dynamics of relevance during the excitation via an ultra-short x-ray pulse, which allows to predict the nuclear inversion.
	
	\subsection{Macroscopic QED Hamiltonian}
	Our starting point is macroscopic QED, a commonly used approach to formulate light-matter interactions in dispersive and absorbing dielectric environments \cite{Scheel2008}. For dielectric media and within the dipole approximation, the resulting Hamiltonian reads \cite{Buhmann2012}
	\begin{align}\label{eq::Dyn_Macro_H}
		\hat{H} =& \int d^3\mathbf{r} \int_{0}^{\infty} d\omega \hbar \omega \hat{\mathbf{f}}^\dagger(\mathbf{r}, \omega) \hat{\mathbf{f}}(\mathbf{r}, \omega) + \sum_{n} \frac{\hbar \omega_{\textrm{nuc}}}{2} \hat{\sigma}^z_{n} \nonumber
		\\
		& - \sum_{n} [\hat{\sigma}^+_{n} \mathbf{d}^* + \hat{\sigma}^-_{n} \mathbf{d}] \cdot \hat{\mathbf{E}}(\mathbf{r}_{n}) \,,
	\end{align}
	where $\hat{\mathbf{f}}(\mathbf{r}, \omega)$ are bosonic operators, $\hat{\mathbf{E}}$ is the electric field operator and polarization is included by vector notation. $\hat{\sigma}^\pm_n$ are the raising/lowering operators associated with nucleus $n$, where we assume for simplicity that only a single transition with dipole moment vector $\mathbf{d}$ is driven (see \cite{LentrodtPhD} for details on how to obtain such a description and the transition parameters for nuclear resonances). The bosonic operators appearing in the Hamiltonian are related to the field operator by \cite{Dung2002}
	\begin{align}\label{eq::Dyn_Macro_EOp}
		\hat{\mathbf{E}}(\mathbf{r}) =& i \sqrt{\frac{\hbar}{\pi \varepsilon_0}}\int_0^\infty d\omega \int d^3\mathbf{r}' \sqrt{\mathrm{Im}[\varepsilon(\mathbf{r}')]} \nonumber \\[2ex]    
		& \qquad \times \mathbf{G}(\mathbf{r}, \mathbf{r}', \omega)\cdot \:\hat{\mathbf{f}}(\mathbf{r}', \omega) \,,
	\end{align}
	where the Green's tensor is defined via
	\begin{align}\label{eq::Dyn_Macro_GreenFn}
		[\nabla\times\nabla\times - \frac{\omega^2}{c^2} \varepsilon(\mathbf{r}, \omega)] \mathbf{G}(\mathbf{r}, \mathbf{r}', \omega) = \delta(\mathbf{r} - \mathbf{r}') \,.
	\end{align}
	This Hamiltonian is a standard tool in optical cavity QED and related fields (see \cite{Scheel2008,Buhmann2012} for reviews). Within the theory of nuclear resonant scattering at x-ray energies, it has previously been used to derive effective level schemes for nuclear ensembles in thin-film cavities ab initio~\cite{Lentrodt2020a}, which are valid for low excitation. In the following, we apply and extend the ideas presented therein to the nonlinear nuclear dynamics during a short x-ray driving pulse.
	
\subsection{Nuclear Master equation}\label{sec::excPh_Master}
The nuclear dynamics are well modeled by the Master equation \cite{Lentrodt2020a}
\begin{align}\label{eq::Dyn_Master_Master}
	\dot{\rho} = -\frac{i}{\hbar}[H_\textrm{eff}, \rho] +  \mathcal{L}_\textrm{eff}[\rho]\,,
\end{align}
where the effective Hamiltonian is given by
\begin{align}\label{eq::Dyn_Master_H}
	\hat{H}_\textrm{eff} =& \sum_{n} \frac{\hbar\omega_{\textrm{nuc}}}{2} \hat{\sigma}_{n}^z - \hbar \sum_{nn'} J_{nn'} \hat{\sigma}_{n}^+ \hat{\sigma}_{n'}^- \nonumber
	\\
	&- \sum_{n} \left[ \textbf{d}^*\cdot\textbf{E}(\mathbf{r}_{n}, t) \hat{\sigma}_{n}^+ + h.c.\right] \,.
\end{align}
The Lindblad term encoding decay processes is
\begin{align}\label{eff::Dyn_Master_L}
	\mathcal{L}_\textrm{eff}[\rho] = &\sum_{nn'} \frac{\Gamma_{nn'}}{2} (2\hat{\sigma}^-_{n}\rho\sigma_{n'}^+ - \{\hat{\sigma}_{n}^+\hat{\sigma}_{n'}^-,\,\rho\}) +\mathcal{L}_\textrm{IC}[\rho]\,,
\end{align}
where $\mathcal{L}_\textrm{IC}[\rho] = \sum_{n} \frac{\gamma_\mathrm{IC}}{2} (2\hat{\sigma}^-_{n}\rho\sigma_{n}^+ - \{\hat{\sigma}_{n}^+\hat{\sigma}_{n}^-,\,\rho\})$ describes the single nucleus loss due to internal conversion. We note that the radiative contribution to the single nucleus decay, often referred to as spontaneous emission rate, is already included in $\Gamma_{nn}$ \cite{Lentrodt2020a}.

The interaction coefficients can be expressed in terms of the Green's function as \cite{Dung2002,Scheel2008,Buhmann2012,Asenjo-Garcia2017a}
\begin{align}
	J_{nn'} \approx \frac{\mu_0 \omega^2_{\textrm{nuc}}}{\hbar} \textbf{d}^*\cdot \mathrm{Re}[\textbf{G}(\mathbf{r}_{n}, \mathbf{r}_{n'}, \omega_{\textrm{nuc},l})]\cdot \textbf{d}
\end{align}
and
\begin{align}
	\Gamma_{nn'} = 2\frac{\mu_0 \omega^2_{\textrm{nuc}}}{\hbar} \textbf{d}^*\cdot \mathrm{Im}[\textbf{G}(\mathbf{r}_{n}, \mathbf{r}_{n'}, \omega_{\textrm{nuc}})]\cdot \textbf{d} \,.
\end{align}
For the layer geometry, the real space Green's function $\textbf{G}(\mathbf{r}_{n}, \mathbf{r}_{n'}, \omega_{\textrm{nuc}})$, which enters these coupling and decay terms, can be obtained via Fourier transforming the analytical result available for the parallel wave space Green's function \cite{Tomas1995,Lentrodt2020a}.

The central approximation used to obtain the above real-space Master equation is the Born-Markov approximation \cite{Dung2002,Lentrodt2020a}. It is a standard tool in light-matter interaction theory \cite{Scully1997,Breuer2002_BOOK} and well applicable at the ultra-weak light-matter coupling of nuclei and x-rays ($\gamma_\mathrm{rad}\sim$~neV $\ll \omega_\mathrm{nuc}\sim$~keV). We emphasize that unlike for the parallel wave space Master equation in the low excitation sector presented in \cite{Lentrodt2020a}, a single parallel wave vector, a low excitation or a linear optics approximation is not employed here. In the above form, the Master equation does also not feature a semi-classical approximation, such that it includes quantum mechanical, collective and cooperative effects \cite{Reitz2022}. \newt{The approximations performed up to this point are standard quantum optics approximations \cite{Scully1997,Carmichael1993,Breuer2002_BOOK,Asenjo-Garcia2017b} and well applicable to the extremely weak coupling regime of narrow x-ray resonances.}

\subsection{The excitation phase}\label{sec::excPhase}

The full quantum driven-dissipative dynamics of a large interacting ensemble of two-level systems governed by Eq.~(\ref{eq::Dyn_Master_Master}) remains a challenging problem of much current interest. However, in the following, we show that the restriction to the excitation phase during the action of a short x-ray pulse considerably simplifies the problem \newt{(see Fig.~\ref{fig::examples_illu_phases} for an illustration)}.

\subsubsection{Time scales and interaction terms}
While the Master equation Eq.~\eqref{eq::Dyn_Master_Master} has a rather simple form, in general, it poses a difficult quantum many-body problem \cite{Gross1982,Agarwal2012}. The cavity environment may mediate interactions between a large number of nuclei, also facilitated by the high coherence of the nuclear ensembles. While it has previously been shown \cite{Lentrodt2020a} that various approximations are applicable in the regime of low excitation, we are here interested in the nonlinear excitation dynamics during an ultra-short pulse. In this case, different approximations can be employed to significantly simplify the equations.

At modern pulsed x-ray sources, such as synchrotrons \cite{Rohlsberger2005} or x-ray free electron lasers (XFELs) \cite{Bostedt2016,Pellegrini2016}, the pulse durations are typically on the picosecond time scale or shorter. For high-intensity XFELs such as the European XFEL \cite{Tschentscher2017}, the pulse length is even on the level of hundreds of femtoseconds if monochromatized. Such pulses were recently employed in first nuclear resonant scattering experiments at  XFELs \cite{Chumakov2018,Shvydko2023}.

In contrast, the natural time scales of the dynamics of M\"ossbauer nuclei are considerably slower~\cite{Hannon1999,Rohlsberger2005}. For example, the M\"ossbauer transition in the archetype isotope ${}^{57}$Fe has a  life time of $141$~ns. Even with superradiant acceleration of the nuclear dynamics, the latter remains orders of magnitude slower than the x-ray pulse duration.

Therefore, in experiments at pulsed x-ray sources, the dynamics  can be divided into two phases. First, in the  \textit{excitation phase}, the nuclei are strongly driven by the ultra-short x-ray pulse. In this short excitation phase, the evolution on natural time scales of the nuclei can safely be neglected. 
Subsequently, in the \textit{decay phase} after the pulse has passed, one observes the decay and collective interaction of the nuclei in the time-dependent emission intensity. The dynamics in this phase occur on the natural time scales of the nuclei, in the absence of external driving fields.

\subsubsection{Excitation phase}

In the following, we are only interested in the maximum excitation of the nuclear ensemble possible with a given x-ray pulse, and therefore can restrict the analysis to the excitation phase. Within this short excitation time, all couplings between the nuclei and and nuclear decay processes  can safely be  neglected, and only the driving term proportional to $\textbf{d}^*\cdot\textbf{E}(\mathbf{r}_{n}, t)$ is of relevance.  Hence, in the excitation phase, the equations of motion resulting from the Master equation Eq.~\eqref{eq::Dyn_Master_Master}  reduce to
\begin{align}
	\frac{d}{dt}{\sigma}^-_{n}  = &-i\omega_{\textrm{nuc}}{\sigma}^-_{n} - i\Omega_{n}\hat{\sigma}^z_{n} \,, \label{eq::eom_SP_sigma_m}
	\\
	\frac{d}{dt}{\sigma}^+_{n}  = &+i\omega_{\textrm{nuc}}{\sigma}^+_{n} + i\Omega^*_{n}{\sigma}^z_{n} \,, \label{eq::eom_SP_sigma_p}
	\\
	\frac{d}{dt}{\sigma}^z_{n}  = & +2i\Omega_{n}{\sigma}^+_{n} - 2i\Omega^*_{n}{\sigma}^-_{n} \,, \label{eq::eom_SP_sigma_z}
\end{align}
where we have defined $\Omega_{n}(t)=\frac{1}{\hbar}\textbf{d}^*\cdot\textbf{E}(\mathbf{r}_{n}, t)$ and operator expectation values $\mathcal{O} = \langle \hat{\mathcal{O}}\rangle$. 

We see that these equations form a closed set already for individual nuclei. We conclude that there are no significant collective effects  in the excitation phase, unlike assumed in earlier works on nuclear excitation by x-ray pulses \cite{Junker2012}. The interactions and collective dynamics only become relevant on the longer nanosecond time scales.

We note that Eqs.~(\ref{eq::eom_SP_sigma_m}--\ref{eq::eom_SP_sigma_z}) are essentially the standard Maxwell-Bloch equations \cite{Scully1997}, which have already been used in a number of works \cite{Liao2011,Junker2012,Liao2012,Liao2013,Kong2014,Kong2016,Liao2016,Wang2018,Zhang2019} on nuclear resonant scattering. 
However, our approach provides a clear approximation hierarchy justifying their validity for nuclei during the excitation phase, for arbitrarily high degrees of excitation. On longer time scales, Eqs.~(\ref{eq::eom_SP_sigma_m}-\ref{eq::eom_SP_sigma_z}) do not necessarily apply and the full Master equation Eq.~\eqref{eq::Dyn_Master_Master} may have to be considered. Our approach therefore resolves discussions surrounding the role of collective effects in the nuclear excitation process \cite{Junker2012,Heeg2016arxiv,Heeg2013b,Lentrodt2020a}.


\subsubsection{Low pulse depletion approximation}\label{sec::approx_lowPulseDepletion}

In addition to the short pulse approximation, we can further simplify the problem by noting that the nuclei's coupling to the radiation field is weak \cite{Rohlsberger2005}. The excitation dynamics can thus be seen as the nuclei simply absorbing photons from the strong driving field, where a significant rate is achieved due to the large number of photons in the pulse. However, the feedback of the nuclear absorption onto the driving pulse can be neglected on the pulse time scale, as we argue in the following.

On the level of a single nucleus, we can consider the Maxwell equation with a nuclear source term
\begin{align}\label{eq::bg_operator_MaxwellBloch_copy}
	\nabla\times\nabla\times {\mathbf{E}}(\mathbf{r}, t) + \frac{\partial^2}{\partial t^2}{\mathbf{E}}(\mathbf{r}, t) = \nonumber
	\\
	\frac{\omega^2_\mathrm{nuc}}{\varepsilon_0\hbar} \sum_n \delta(\mathbf{r}-\mathbf{r}_{n}) [\mathbf{d}^*\hat{\sigma}_n^+(t)  + \mathbf{d} \hat{\sigma}_n^-(t) ] \,,
\end{align}
which couples to the optical Bloch equations Eqs.~(\ref{eq::eom_SP_sigma_m}-\ref{eq::eom_SP_sigma_z}). We then see that while the driving terms in the latter are of first order in $\mathbf{d}$, the source term in the Maxwell equation is of second order. Consequently, the fractional number of absorbed photons is typically small, such that the ultra-short pulse field is not significantly changed by the nuclear absorption. Note that on larger time scales during the decay phase, when there is no driving pulse, the nuclear emission then constitutes the leading order and does have to be accounted for. 

As a result, we can usually approximate the driving term in the optical Bloch equation as
\begin{align}
	\Omega_{n}(t)=\frac{1}{\hbar}\textbf{d}^*\cdot\textbf{E}(\mathbf{r}_{n}, t)\approx\frac{1}{\hbar}\textbf{d}^*\cdot\textbf{E}_\mathrm{drive}(\mathbf{r}_{n}, t) \,,
\end{align}
where $\textbf{E}_\mathrm{drive}$ is the driving field in absence of the nuclear resonances. This approximation is particularly useful, since it uncouples the optical Bloch equations from the Maxwell equation.

While this argument is fully correct for a single nucleus, we have to further consider the effect of adding many nuclei. For high nuclear number density or large propagation depths, the sum over the nuclei in Eq.~\eqref{eq::bg_operator_MaxwellBloch_copy} can potentially result in significant pulse depletion. In particular for thick target forward scattering, such effects may have to be taken into account \cite{Hoy1999,Rohlsberger2005}. We note, however, that in such cases, the approximation results in an overestimation of the driving field. Consequently, to achieve maximum inversion throughout the ensemble, it is experimentally advantageous to stay within the validity region of this approximation by using optically thin resonant ensembles.

In the context of propagation problems such as thick target forward scattering, the well-known effect of multiple resonant scattering \cite{Lynch1960,Smirnov1999} is effectively also precluded by this approximation. We note, however, that multiple scattering mainly occurs on longer time scales during the decay phase, which the above approximation does not apply to in the first place.

We will utilize this approximation of neglecting resonant feedback for the calculations in the following. Its validity is confirmed in Appendix~\ref{app:lowPulseDepConfirm}for the samples considered in the companion paper \cite{companion_nonlin2024_arxivLetter}.

\section{Characterizing the  excitation dynamics  via the pulse area theorem}\label{sec::excDyn_pulseArea}
In the following, we discuss an analytical solution to the above optical Bloch equations  known as the pulse area theorem~\cite{McCall1967,Allen1987_BOOK,Eberly1998,Shore2011,Fischer2017}. For our purpose, this analytical solution is particularly useful since in combination with the semi-analytical approach to pulse propagation through the cavity developed in Sec.~\ref{sec::exPh_beamDiv}, it tremendously simplifies the numerical complexity of the nuclear excitation calculation (see Fig.~\ref{fig::examples_illu_algo}).

The pulse area theorem applies under certain conditions~\cite{Shore2011,Fischer2017}. Some of them, such as the absence of interaction and spontaneous emission terms, are already ensured by the time scale separation during the driving phase discussed above. The main condition of relevancce is that of  a coherent resonant driving field. Specifically, if the driving field is of the form
\begin{align}
	\Omega_{n}(t) = |\Omega_{n}(t)| e^{i\phi_n}e^{-i\omega_{\textrm{nuc}}t}
\end{align}
where $\phi_n$ is a real time-independent constant, then the solution for the nuclear expectation values is \cite{Shore2011}
\begin{align}
	{\sigma}^z_{n}(t) &= -\cos(\Phi_{n}(t)) \,, \label{eq::pulse_area_solz_phi}
	\\
	{\sigma}^-_{n}(t) &= +\frac{i}{2}e^{-i\omega_{\textrm{nuc}}t}e^{i\phi_{n}}\sin(\Phi_{n}(t)) \,, \label{eq::pulse_area_solm_phi}
	\\
	{\sigma}^+_{n}(t) &= -\frac{i}{2}e^{i\omega_{\textrm{nuc}}t}e^{-i\phi_{n}}\sin(\Phi_{n}(t)) \,, \label{eq::pulse_area_solp_phi}
\end{align}
where $\Phi_{n}(t)$ is the pulse area \cite{McCall1967,Shore2011} defined by
\begin{align}
	\Phi_{n}(t) = \int_{t_0}^t dt' 2|\Omega_{n}(t')| \,,
\end{align}
and $t_0$ is the initial time, at which we assume the nuclei to be in their ground state with ${\sigma}^z_{n} = -1$, ${\sigma}^\pm_{n}=0$. The latter condition is typically well valid at room temperature in the absence of hyperfine splittings \cite{Rohlsberger2005,HeegPhD}.

The solution applies if the carrier frequency of the driving field is resonant with the nuclear transition and the envelope has a constant phase, and therefore assumes an idealized highly monochromatized seeded x-ray pulse. For the preset analysis, this is not a crucial restriction, since highest resonant flux together with the removal of the dominant spectral parts off-resonance with the nuclei using high-resolution monochromators is a key requirement for any experimental implementation of the high-excitation scenario, in order to reduce radiation damage to the sample.

We note that in cases where  the pulse area theorem does not apply, the optical Bloch equations can straightforwardly be solved numerically. However, the pulse area theorem has a convenient relation to frequency domain response functions which will be useful later on. In particular, if the pulse arrives from a large distance and if we are only interested in the final excitation after the whole pulse has passed, the relevant limit of the pulse area is $t_0\to-\infty$ and $\Phi^{\mathrm{tot}}_n = \lim\limits_{t \to \infty} \Phi_{n}(t)$. The pulse area then directly relates to the frequency domain amplitude on resonance, such that
\begin{align}\label{eq::pulse_area_tot}
	\Phi^{\mathrm{tot}}_{n} = \frac{4\pi}{\hbar}|\textbf{d}^*\cdot\textbf{E}_\mathrm{drive}(\mathbf{r}_{n}, \omega_\mathrm{nuc})| \,,
\end{align}
where we define time- to frequency-domain Fourier transforms according to
\begin{align}
	\textbf{E}(\mathbf{r}_{n}, \omega) = \frac{1}{2\pi}\int dt e^{i\omega t} \textbf{E}(\mathbf{r}_{n}, t) \,.
\end{align}
We will use this property in Sec.~\ref{sec::exPh_beamDiv} for a semi-analytical description of nuclear excitation in thin-film cavities by focused pulses without the need for additional approximations.

\section{Focussed beam driving in the thin-film geometry}\label{sec::exPh_beamDiv}

According to Eq.~\eqref{eq::pulse_area_tot}, the central property of the light field which determines the excitation after the pulse is the Fourier amplitude of the field $\textbf{E}_\mathrm{drive}(\mathbf{r}_{n}, \omega_\mathrm{nuc})$ at the location of the nuclei. When the narrow resonances are placed inside a photonic environment such as a thin-film cavity~\cite{Heeg2013b,Lentrodt2020a}, the input field provided by the x-ray source is further modified by off-resonant electronic refraction, absorption and the resulting interference, which influence the driving field at the location of the nuclei. In the case of a highly collimated beam, the resulting cavity response is well known from Parratt's formalism~\cite{Parratt1954} or the related transfer matrix formalism~\cite{Abeles1950,Rohlsberger2005,LentrodtPhD}, and software packages for its application to nuclear resonance scattering~\cite{Sturhahn2000,Shvydko2000,pynuss} as well as to simple model systems \cite{Lentrodt2022_pyrot} are available.

In this section, we develop a semi-analytical formalism to compute nuclear inversion inside such thin-film x-ray cavities under focused x-ray driving. A key challenge arises from the fact that the focused beam itself is most conveniently characterized in a coordinate system with axes in the direction of propagation and in the plane perpendicular to it. The field inside the cavity, however, is most conveniently described using coordinates axes in the cavity surface plane and along its normal. Therefore,  the electric field of the pulse needs to be transformed between the two coordinate systems.

Using the approximate results from previous sections, we first derive expressions for the free space field's Fourier amplitude in terms of common beam parameters of x-ray sources, which then allow to predict nuclear excitation fractions. Next, we introduce a scheme based on Fourier transforms to include the cavity response for focused x-ray pulses beyond the highly collimated case. The approach relies on the analytical insights from previous sections to reduce the number of necessary numerical Fourier transforms.

\subsection{Pulsed Gaussian beam in free space}\label{sec::cav_freeGauss}

\subsubsection{Angular spectrum and real space field}
Assuming transverse coherence, the x-ray pulse  can  be represented as a pulsed version of the archetype Gaussian beam \cite{Born1980,Siegman1986,Mandel1995}. The latter can most concisely be represented by its angular spectrum defined by \cite{NovotnyHecht2006}
\begin{align}
	\tilde{\textbf{E}}(\mathbf{k}^\mathrm{p}_{\perp}, z_\mathrm{p}, \omega) = \frac{1}{4\pi^2} \int d^2 \mathbf{r}_\perp e^{-i\mathbf{k}^\mathrm{p}_{\perp}\cdot\mathbf{r}^\mathrm{p}_{\perp}} \tilde{\textbf{E}}(\mathbf{r}_\mathrm{p}, \omega) \,,
\end{align}
where $z_\mathrm{p}$ is the spatial coordinate along the beam propagation direction, $\mathbf{r}_{\perp}$ the vector in the perpendicular plane and $\mathbf{k}_\perp$ its corresponding wave vector. The sub- or superscripts ``p'' (``c'') indicate pulse (cavity) coordinates [see Fig.~\ref{fig::examples_illu_setup}] and we use the complex phasor notation for the electric field, which is indicated by the tilde.

The Gaussian beam is given by the solution \cite{NovotnyHecht2006}
\begin{align}\label{eq::GaussBeam_angSpec}
	\tilde{\textbf{E}}^\mathrm{Gauss}(\mathbf{k}^\mathrm{p}_{\perp}, z_\mathrm{p}, \omega) = \mathbf{A}(\omega) \frac{w_0^2}{2} e^{-\frac{w_0^2}{4}|\mathbf{k}^\mathrm{p}_\perp|^2} e^{i z_\mathrm{p} k^{\mathrm{p},\mathrm{val}}_z}\,,
\end{align}
where $w_0$ is the beam waist size at the focus and $k^{\mathrm{p},\mathrm{val}}_z$ is a function of the frequency and angular wave vector given by $k^{\mathrm{p},\mathrm{val}}_z = \sqrt{k^2 - |\mathbf{k}^\mathrm{p}_\perp|^2}$. The beam waist size is further related to the beam divergence $\theta_\mathrm{div}$ by $w_0 = \frac{2}{k\theta_\mathrm{div}}$, where we assume full spatial coherence. $\mathbf{A}(\omega)$ is the frequency spectrum of the pulse.

The real-space field can then be obtained by a Fourier transform of the angular spectrum in the perpendicular plane. For the case of a paraxial beam, the real space field can be approximated analytically as \cite{NovotnyHecht2006}
\begin{align}\label{eq::cav_freeGauss_realSpace_analytic}
	\tilde{\textbf{E}}^\mathrm{Gauss}(\mathbf{r}_\mathrm{p}, \omega) \approx& 2\pi \mathbf{A}(\omega) \frac{w_0}{w(z_\mathrm{p})}e^{ikz_\mathrm{p}} e^{-\frac{|\mathbf{r}^\mathrm{p}_\perp|^2}{w^2(z_\mathrm{p})}} \nonumber
	\\
	&\times e^{-i\psi(z_\mathrm{p})} e^{ik\frac{|\mathbf{r}^\mathrm{p}_\perp|^2}{2R(z_\mathrm{p})}} \,,
\end{align}
where the Gaussian beam parameters are the beam size $w(z) = w_0\sqrt{1+\frac{z^2}{z_R^2}}$, the Gouy phase $\psi(z)=\arctan(\frac{z}{z_R})$, the phase curvature radius $R(z) = z(1+\frac{z_R^2}{z^2})$ and the Rayleigh length $z_R=\frac{kw_0^2}{2}$.

For a pulse whose frequency spectrum is narrow compared to the wavelength, we can further assume that only the carrier frequency phase varies significantly with energy. We can then obtain a simple form of a pulsed Gaussian beam in the space and time domain as
\begin{align}
	\tilde{\textbf{E}}^\mathrm{Gauss}(\mathbf{r}_\mathrm{p}, t)=& 2\pi {\mathbf{A}}(t - z_\mathrm{p}/c) \frac{w_0}{w(z)}e^{ikz_\mathrm{p}} e^{-\frac{|\mathbf{r}^\mathrm{p}_\perp|^2}{w^2(z_\mathrm{p})}} \nonumber
	\\
	&\times e^{-i\psi(z_\mathrm{p})} e^{ik\frac{|\mathbf{r}^\mathrm{p}_\perp|^2}{2R(z_\mathrm{p})}} \,,
\end{align}
where $\tilde{\textbf{E}}(\mathbf{r}, t) = \int d\omega e^{-i\omega t}\tilde{\textbf{E}}(\mathbf{r}, \omega)$ and $\mathbf{A}(t) = \int d\omega e^{-i\omega t}\mathbf{A}(\omega)$. We note that due to the complex phasor notation, the physical electric field is given by $\textbf{E}(\mathbf{r}, t) = \mathrm{Re}[\tilde{\textbf{E}}(\mathbf{r}, t)]$.

\subsubsection{Amplitude normalization}\label{sec::amp_norm}
For a Fourier limited pulse, the time domain spectrum has the form $\mathbf{A}(t) = \frac{\mathbf{A}_0}{\tau_\mathrm{pulse}\sqrt{2\pi}} e^{-\frac{t^2}{2\tau_\mathrm{pulse}^2}}e^{-i\omega_{\textrm{nuc}}t}$,
such that $\mathbf{A}(\omega) =\frac{\mathbf{A}_0}{2\pi} e^{-\frac{\tau_\mathrm{pulse}^2}{2}(\omega-\omega_{\textrm{nuc}})^2}$, where we have chosen the normalization such that $\mathbf{A}_0$ is the peak amplitude of $\tilde{\mathbf{E}}^\mathrm{Gauss}(\mathbf{r}_\mathrm{p}, \omega)$.

The amplitude can be fixed via the pulse energy or photon number
\begin{align}
	N_\mathrm{ph} \hbar \omega_\mathrm{nuc} \approx \frac{\varepsilon_0}{2} \int d^3\mathbf{r} |\tilde{\textbf{E}}(\mathbf{r}, t)|^2 \,,
\end{align}
where we have again assumed a narrow spectrum compared to the wavelength and a much larger focus size than the carrier wavelength.

Evaluating the integral for the pulsed Gaussian beam analytically gives
\begin{align}
	|\mathbf{A}_0| = \sqrt{\frac{2 N_\mathrm{ph}\hbar \omega_\mathrm{nuc} \tau_\mathrm{pulse}}{\pi^2\sqrt{\pi}\varepsilon_0 w_0^2 c }} \,.
\end{align}
Within the rotating wave approximation \cite{Scully1997}, which in this context can be phrased as $\Omega_n(t) = \frac{1}{\hbar}|\textbf{d}^*\cdot\textbf{E}_\mathrm{drive}(\mathbf{r}_{n}, t)|\approx \frac{1}{2\hbar}|\textbf{d}^*\cdot\tilde{\textbf{E}}_\mathrm{drive}(\mathbf{r}_{n}, t)|$, the pulse area theorem from Sec.~\ref{sec::excDyn_pulseArea} applies to this pulse and can be used to solve the optical Bloch equations. Using Eq.~\eqref{eq::pulse_area_tot} for aligned polarization and dipole transition direction, the peak of the full pulse area is then given by
\begin{align}\label{eq::peakPulseArea}
	\Phi^{\mathrm{tot}}_\mathrm{peak} \approx \frac{2\pi}{\hbar} |\mathbf{d} \cdot \mathbf{A}_0| =  \sqrt{\frac{8 N_\mathrm{ph} \omega_\mathrm{nuc} \tau_\mathrm{pulse} d^2}{\sqrt{\pi} w_0^2 c\hbar \varepsilon_0}}\,.
\end{align}

This formula is key in obtaining an absolute scale for the inversion when the parameters of the x-ray source beam are given. Full inversion of the nuclei at the peak intensity location can be realized when $\Phi^{\mathrm{tot}}_\mathrm{peak}>\pi$ \cite{Heeg2016arxiv}, since $\sigma^z=1$ for $\Phi=\pi$ according to Eq.~\eqref{eq::pulse_area_solz_phi}.

\subsubsection{Intuitive decomposition of the pulse area}
Instead of $\omega_\mathrm{nuc}, N_\mathrm{ph}, \tau_\mathrm{pulse}$, a better set of characteristic variables for the incoming pulse is $\omega_\mathrm{nuc}$, $E_\mathrm{pulse}, b_\mathrm{r}$, where $E_\mathrm{pulse}$ is the pulse energy and $b_\mathrm{r}$ is the relative bandwidth of the pulse. For a Fourier limited pulse, we have
\begin{align}
	\tau_\mathrm{FWHM} \Delta \nu = \frac{2 \ln(2)}{\pi} \approx 0.441
\end{align}
with $\tau_\mathrm{FWHM}=\sqrt{4\ln(2)}\,\tau_\mathrm{pulse}\approx 1.67\,\tau_\mathrm{pulse}$. With $\Delta \nu = b_r \nu = b_r \omega_\mathrm{nuc}/(2\pi)$, we then conclude that
\begin{align}
	\tau_\mathrm{pulse} = \frac{\sqrt{\ln(2)}}{b_r \omega_\mathrm{nuc}} \,.
\end{align}
For the pulse energy, we already used above that
\begin{align}
	N_\mathrm{ph} = \frac{E_\mathrm{pulse}}{\hbar \omega_0} \,.
\end{align}
The pulse area can then also be expressed as
\begin{align}\label{eq::peakPulseArea2}
	\Phi^{\mathrm{tot}}_\mathrm{peak} \approx \frac{2\pi}{\hbar} |\mathbf{d} \cdot \mathbf{A}_0| =&  \sqrt{\frac{8 \sqrt{\ln(2)} E_\mathrm{pulse} d^2}{\sqrt{\pi} w_0^2 c\hbar \varepsilon_0 \hbar \omega_\mathrm{nuc} b_r}} \nonumber
	\\
	\approx& 1.9385 \cdot \frac{\sqrt{\tilde{\sigma}_\mathrm{nuc}}}{w_0} \cdot \sqrt{\frac{E_\mathrm{pulse}}{b_r\,\hbar \omega_\mathrm{nuc}}}\,,
\end{align}
where $\sqrt{\tilde{\sigma}_\mathrm{nuc}}:=\frac{d}{\sqrt{c\hbar \varepsilon_0}}$, which has dimensions of length. This formula can also be understood intuitively. The first term is essentially the ratio between the root of a cross-section and the root of the beam area, while the second term is the root of the number of photons within the bandwidth.

Note that while we derived this formula using the pulse area theorem and other assumptions which only apply for a transform-limited pulse, the final result is likely more broadly applicable, since we use the bandwidth instead of the pulse duration. That is, Eq.~\eqref{eq::peakPulseArea2} likely generalizes (at least approximately) beyond the Fourier-limited case while Eq.~\eqref{eq::peakPulseArea} does not. Strictly speaking, the equations are, however, derived for a Fourier-limited pulse.

Note also that in the case of M\"ossbauer transitions, $\tilde{\sigma}_\mathrm{nuc}$ is not equal to what is commonly referred to as the nuclear absorption cross-section \cite{Rohlsberger2005}, which can be expressed as
\begin{align}
	\sigma_0 = \frac{2\pi}{(1+\alpha)k_0^2} \frac{2I_e + 1}{2I_g + 1}\,,
\end{align}
where $k_0$ is the wavenumber corresponding to $\omega_\mathrm{nuc}$. Written out, we instead have
\begin{align}
	\tilde{\sigma}_\mathrm{nuc} = \frac{2\pi}{(1+\alpha)k_0^2} \frac{2I_e + 1}{2I_g + 1} \frac{\gamma}{2\omega_\mathrm{nuc}}\,,
\end{align}
where $\gamma$ is the linewidth of the nuclear transition (e.g.~$\gamma=4.66\,$neV for $^{57}$Fe) and $\alpha$ is the internal conversion coefficient (e.g.~$\alpha=8.6$ for $^{57}$Fe)~\cite{Rohlsberger2005}. $\sigma_0$ and $\tilde{\sigma}_\mathrm{nuc}$ are therefore proportional to each other, but $\tilde{\sigma}_\mathrm{nuc}$ crucially includes information about the nuclear linewidth and therefore about the dipole moment. For general transitions, such as $M1$ transitions which are found in many Mössbauer nuclei, one has to be careful to compute an appropriate effective dipole moment \cite{LentrodtPhD} or switch to magnetic fields~\cite{Andrejic2021} in the coupling Hamiltonian.

We can then define two characteristic dimensionless quantities. The first is independent of the bare source parameters (before focusing) and defines the nuclear and focusing properties
\begin{align}\label{eq::chi_sigma}
	\chi_\sigma \approx 1.9385 \cdot \sqrt{\frac{\tilde{\sigma}_\mathrm{nuc}}{\hbar \omega_0}} \,.
\end{align}
The second quantity is essentially the number of photons within the bandwidth
\begin{align}
	\chi_\mathrm{source}:=\sqrt{\frac{E_\mathrm{pulse}}{b_r}}= \sqrt{\frac{\hbar \omega_0 N_\mathrm{ph} \, \omega_0 \tau_\mathrm{pulse}}{\sqrt{\ln(2)}}}\,.\label{eq::chi_source_def}
\end{align}
The pulse area can then be decomposed as
\begin{align}
	\Phi^{\mathrm{tot}}_\mathrm{peak} = \frac{\chi_\sigma}{w_0} \cdot \chi_\mathrm{source} \,. \label{eq::pulse_area_neat}
\end{align}
This decomposition allows to define the intuitive quantity
\begin{align}
	{\chi}_\mathrm{source}^\mathrm{nec}(w_0) = \frac{\pi w_0}{{\chi}_\sigma} \,,
\end{align}
which quantifies the \textit{necessary} source characteristic to reach inversion. That is when ${\chi}_\mathrm{source}^\mathrm{nec}(w_0) = \chi_\mathrm{source}$, one reaches inversion.

In a companion paper \cite{companion_nonlin2024_arxivLetter}, we investigate this quantity for different M\"ossbauer isotopes in order to provide a general idea of what can be expected at different sources. This decomposition further allows to easily visualize the expected nuclear excitation for various scenarios. Note that there are different reasonable choices of decomposition. The above choice, in particular including the $\hbar \omega_0$ factor in Eq.~\eqref{eq::chi_sigma}, is motivated by the fact that $E_\mathrm{pulse}$ and $b_r$ typically only weakly depend on the pulse energy at XFEL sources \cite{Madsen2013_designRep,Madsen2021}, such that $\chi_\mathrm{source}$ is roughly constant over a relatively wide parameter range.

\subsection{Cavity field enhancement}

\subsubsection{Cavity response}
When the nuclei are embedded in a cavity, the free space field from the previous section is modified by reflections from and absorption in the cavity material. For a highly collimated monochromatic field, the field distribution can be calculated straightforwardly using the standard Parratt's formalism \cite{Parratt1954} or the extended layer formalism (see e.g.~\cite{Rohlsberger2005,LentrodtPhD} and references therein). The effect of the cavity is then given by the mode functions \cite{Tomas1995} $\mathbfcal{E}^{0(n)}_{q/p}(z_\mathrm{c}, \mathbf{k}^\mathrm{c}_\parallel, \omega)$ for polarization $q/p$, depth from the upper cavity boundary $z$, and the wave vector in the cavity plane $\mathbf{k}^\mathrm{c}_\parallel$. $0$ $(n)$ signifies an incident field from the top (bottom) of the cavity.

These cavity response functions are available analytically for the planar thin-film geometry~\cite{Tomas1995,Lentrodt2020a}, providing a useful basis to expand the field in the form
\begin{align}
	\tilde{\mathbf{E}}(\mathbf{r}_\mathrm{c}, \omega) = \int d^2\mathbf{k}^\mathrm{c}_\parallel e^{i\mathbf{k}^\mathrm{c}_\parallel \cdot \mathbf{r}^\mathrm{c}_\parallel} \mathbfcal{E}^{0}_s(z_\mathrm{c}, \mathbf{k}^\mathrm{c}_\parallel, \omega) E_\textrm{in}(\mathbf{k}^\mathrm{c}_\parallel, \omega) \,, \label{eq::cav_resp_RespFunsExp}
\end{align}
where we consider the special case of a purely $s$-polarized field incident from the top of the cavity for simplicity. $E_\mathrm{in}(\mathbf{k}_\parallel, \omega)$ are appropriate Fourier coefficients of the incoming field from the x-ray source, which appear in a similar expansion of a field in free space without a cavity, given by
\begin{align}
	\tilde{\textbf{E}}_\mathrm{in}(\mathbf{r}_\mathrm{c}, \omega) = \int d^2\mathbf{k}^\mathrm{c}_\parallel  e^{i\mathbf{k}^\mathrm{c}_\parallel\cdot\mathbf{r}^\mathrm{c}_\parallel}e^{ik^{\mathrm{c},\mathrm{val}}_z z_\mathrm{c}} E_\textrm{in}(\mathbf{k}^\mathrm{c}_\parallel, \omega) \hat{\mathbf{e}}_s  \,,
\end{align}
where $k^{\mathrm{c},\mathrm{val}}_z = \sqrt{k^2 - |\mathbf{k}^\mathrm{c}_\parallel|^2}$. Inverting the Fourier relation, we can obtain the mode coefficients from the input field as
\begin{align}\label{eq::cav_resp_Ein_FT}
	E_\textrm{in}(\mathbf{k}_\parallel, \omega) = \frac{e^{-ik^{\mathrm{c},\mathrm{val}}_z z_\mathrm{c}}}{2\pi}\int d^2\mathbf{r}^\mathrm{c}_\parallel  e^{-i\mathbf{k}^\mathrm{c}_\parallel\cdot\mathbf{r}^\mathrm{c}_\parallel} \hat{\mathbf{e}}_s \cdot \tilde{\textbf{E}}_\mathrm{in}(\mathbf{r}_\mathrm{c}, \omega) \,.
\end{align}
Note that the $z_\mathrm{c}$-dependence here exactly cancels, such that $E_\mathrm{in}$ is independent of $z_\mathrm{c}$

\subsubsection{Fourier coefficients of the incoming field}
Noting that the expansion in terms of cavity mode functions Eq.~\eqref{eq::cav_resp_RespFunsExp} is an angular spectrum in rotated coordinates, the Fourier coefficients $E_\textrm{in}(\mathbf{k}_\parallel, \omega)$ can indeed be obtained analytically for the Gaussian beam considered in Sec.~\ref{sec::cav_freeGauss}. The result is derived in Appendix \ref{app::gaussCavCoord} and reads
\begin{widetext}
	\begin{align}\label{eq::cav_cavAngSpec_analytic}
		E^\mathrm{Gauss}_\textrm{in}(\mathbf{k}^\mathrm{c}_\parallel, \omega)  = \frac{\hat{\mathbf{e}}_s \cdot\tilde{\mathbf{A}}(\omega)}{2\pi} \tilde{I}\left(k_x, k_y, \sqrt{k^2 - k_y^2 - k_x^2} \right) F_\delta\left(k_x, k_y, \sqrt{k^2 - k_y^2 - k_x^2} \right)\Theta(k^2 - k_y^2 - k_x^2) \,,
	\end{align}
which we refer to as the cavity angular spectrum of the incoming field in the following. Here, we define
\begin{align}
	\tilde{I}(k_x, k_y, k_z) &= \frac{w_0^2}{2} e^{-\frac{w_0^2}{4} \left[k_y^2 + k_x^2\sin^2(\theta_\mathrm{in}) + k_z^2\cos^2(\theta_\mathrm{in}) - 2k_x k_z \cos(\theta_\mathrm{in}) \sin(\theta_\mathrm{in}) \right]}\,,
\end{align}
\end{widetext}
\begin{align}
	F_\delta\left(k_x, k_y, k_z\right)= \left|\sin(\theta_\mathrm{in}) +\frac{k_x\cos(\theta_\mathrm{in})}{k_z}\right| \,,
\end{align}
and $\theta_\mathrm{in}$ is the incidence angle between the Gaussian beam's propagation axis and the cavity plane. We further use $\mathbf{k}^\mathrm{c}_\parallel = (k_x, k_y)$, $\mathbf{r}^\mathrm{c}_\parallel = (x, y)$ and $z_c=z$ as an abbreviated notation in the following. For an illustration of the geometry and coordinate systems, see Fig.~\ref{fig::examples_illu_setup}.

\begin{figure}[t]
	\centering
	\includegraphics[width=1.0\columnwidth]{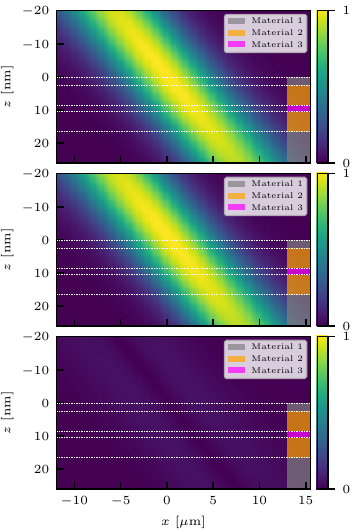}
	\caption{Comparison of resonant field intensities $|\mathbf{E}(\mathbf{r}, \omega)|^2$ at $y=0$ in free space (realised by taking the limit of unity refractive index for the cavity structure indicated by white dashed lines and color shading on the right) for a focused beam with $\theta_\mathrm{div}=1.1~$mrad, $\theta_\mathrm{in}\approx3.352~$mrad. Top: The analytical formula for the Gaussian beam in the paraxial approximation Eq.~\eqref{eq::cav_freeGauss_realSpace_analytic}. Middle: The result of the numerical Fourier transform of the cavity angular spectrum Eq.~\eqref{eq::cav_cavAngSpec_analytic}. Bottom: The intensity of the difference of the two fields.}
	\label{fig::examples_freeSpaceLimit}
\end{figure}

\subsubsection{Numerical approach}
Once the cavity angular spectrum of the incoming field is known, the field in real space can be obtained by numerically evaluating Eq.~\eqref{eq::cav_resp_RespFunsExp}. To this end, we choose a grid in $\mathbf{k}^\mathrm{c}_\parallel$-space and use the \textsc{scipy} \cite{scipy2020} fast Fourier transform algorithms to perform the integrals.

In our approach here and in the companion paper~\cite{companion_nonlin2024_arxivLetter}, we use the analytical solution for the Gaussian beam's angular cavity spectrum Eq.~\eqref{eq::cav_cavAngSpec_analytic} as an input to the algorithm. For more general spatial beam profiles, the Fourier transform Eq.~\eqref{eq::cav_resp_Ein_FT} can, however, be performed analogously by numerical means.

For convenience, we then define the cavity field enhancement factor $\xi_\mathcal{E}$ as the ratio of the field amplitude in the cavity at the resonant layer's center for a given focusing to the peak of the field amplitude in free space. We can then use the free space result for the nuclear inversion in form of the pulse area Eq.~\eqref{eq::peakPulseArea} multiplied by the $\xi_\mathcal{E}$, which can be calculated by the above field propagation formalism for an appropriately normalized input pulse.

\subsection{Benchmarks and illustration}\label{sec::beamDiv_ex}
To test and illustrate our approach, we consider an exemplary thin-film x-ray cavity, which is doped with an ensemble of M\"ossbauer nuclei at its center. The cavity structure is Pt (2.5 nm)/C (6 nm)/$^{57}$Fe (2 nm)/C (6 nm)/Pt, such that the first mode is close to critically coupled. The incidence angle is chosen at the first resonance minimum $\theta_\mathrm{in} = \theta_\mathrm{min,1} \approx 3.352~$mrad.

\subsubsection{No cavity limit}
In the free space limit, that is setting all refractive indices of the cavity layers to unity to mimic no cavity at all, we can benchmark the numerical Fourier transform by comparison to the analytical formula in the paraxial approximation given by Eq.~\eqref{eq::cav_freeGauss_realSpace_analytic}. Results for a beam divergence of $\theta_\mathrm{div}=1.1~$mrad are shown in Fig.~\ref{fig::examples_freeSpaceLimit}, demonstrating excellent agreement. The remaining deviations are due to the finite Fourier grid. The figure shows the normalized beam intensity profile in the $x$-$z$-plane at $y=0$, which constitutes an excerpt of the expected Gaussian beam profile. 

\subsubsection{Highly collimated limit}
\begin{figure}[t]
	\centering
	\includegraphics[width=1.0\columnwidth]{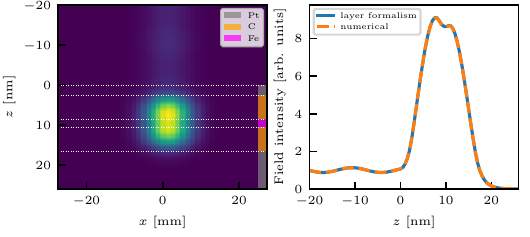}
	\caption{\newt{Highly collimated limit ($\theta_\mathrm{div}=1~\mu$rad) of a focused x-ray beam incident on the cavity. Left: Resonant field intensity $|\mathbf{E}(\mathbf{r}, \omega)|^2$ at $y=0$. In this limit, the cavity enhancement does not depend on $x$ and the profile in the $z$-direction is proportional to the result calculated in the analytical layer formalism \cite{Rohlsberger2005}, which assumes a plane wave input. Right: Slice of the resonant field intensity at $x=y=0$, to cross-check this feature. The layer formalism result (solid blue line) matches our numerical algorithm in this limit (dashed orange line). The amplitude normalization was chosen such that the peaks of the two calculations coincide. Note the difference in scale on the $x$-axis compared to Fig.~\ref{fig::examples_freeSpaceLimit} due to the low beam divergence here.}}
	\label{fig::examples_lowDivLimit}
\end{figure}
\begin{figure}[t]
	\centering
	\includegraphics[width=1.0\columnwidth]{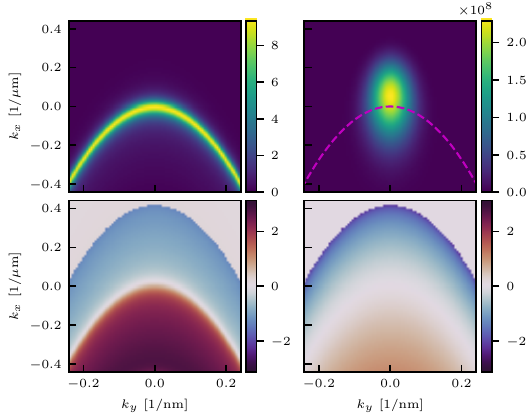}
	\caption{Illustration of the two basic quantities [beam mode coefficients and mode functions; see Fig.~\ref{fig::examples_illu_algo}] used in the Fourier transformation algorithm for the case of the example cavity. The angular spectra are evaluate at the $z$-position of the nuclear ensemble center. (left; top and bottom) Intensity and phase, respectively, of the cavity response function $\mathbfcal{E}^{0}_s(z, \mathbf{k}^\mathrm{c}_\parallel, \omega)$. (right; top and bottom) Intensity and phase, respectively, of the angular spectrum in cavity coordinates $E_\textrm{in}(\mathbf{k}^\mathrm{c}_\parallel, \omega)$ of a normalized focused Gaussian input pulse (carrier phases omitted) with $\theta_\mathrm{in} \approx 3.352~$mrad. The dashed magenta line indicates the angular contour corresponding to the incidence angle.  Interestingly, this line is shifted away from the peak of the cavity angular spectrum of the pulse, despite the pulse being centered on the minimum in bare spatial coordinates. The parabola above which the phases are zero corresponds to $k_z^{(\mathrm{val})}=0$.}
	\label{fig::examples_basicQuantities}
\end{figure}
As a second benchmark, we can compare to the well-studied case of low focusing \cite{Rohlsberger2005}. For this case of highly collimated beams, the mode functions $\mathbfcal{E}^{0}_s(z_\mathrm{c}, \mathbf{k}^\mathrm{c}_\parallel, \omega)$ are approximately constant over the relevant range of parallel wave vectors set by $E_\textrm{in}(\mathbf{k}^\mathrm{c}_\parallel, \omega)$. One can then factor out the mode functions from the integral, such that
\begin{align}\label{eq::cav_resp_RespFunsExp_collimated}
	\tilde{\mathbf{E}}(\mathbf{r}_\mathrm{c}, \omega) \approx \mathbfcal{E}^{0}_s(z_\mathrm{c}, \mathbf{k}^{\mathrm{c},(\mathrm{in})}_\parallel, \omega) \int d^2\mathbf{k}^\mathrm{c}_\parallel e^{i\mathbf{k}^\mathrm{c}_\parallel \cdot \mathbf{r}^\mathrm{c}_\parallel} E_\textrm{in}(\mathbf{k}^\mathrm{c}_\parallel, \omega) \,.
\end{align}
The result is thus an envelope in the $(x,y)$-plane parallel to the cavity with a constant enhancement factor independent of the small beam divergence. This value should further correspond to the field intensity profile that can be calculated in the so-called layer formalism \cite{Rohlsberger2005,Sturhahn2000,pynuss}, which computes the interference for an incident plane wave --- a perfectly collimated beam.

As a test, Fig.~\ref{fig::examples_lowDivLimit} illustrates the results of our algorithm in this limit for the example cavity. Comparison to the layer formalism \cite{Rohlsberger2005,pynuss} shows that our algorithm yields the correct result in this limit.

\subsubsection{Cavity response and beam angular spectra}
To provide intuition on the working of the approach, Fig.~\ref{fig::examples_basicQuantities} shows the two central quantities --- both of which are computed analytically --- which enter the numerical Fourier transform. These are the cavity response function and the beam angular spectrum in cavity coordinates (see Fig.~\ref{fig::examples_illu_algo}). In Fig.~\ref{fig::examples_basicQuantities}, both of them are shown at a fixed $z$ which is chosen as the center of the resonant ensemble.

We see that the cavity has a peaked response due to a dominant guiding mode, whose incidence angle is fixed such that its central $k_x$ varies with $k_y$. The contour can be understood by noting that the resonance condition for a mode is determined by $k_z$ and therefore via the incidence angle due to $k_z=k \cos(\theta_\mathrm{in})$, such that $k^2 = k_x^2 + k_y^2 + k_z^2$ leads to $k_x = \sqrt{k^2 \sin^2(\theta_\mathrm{in}) - k_y^2}$. The mode profile is then a resonance peak around this contour.

For completeness, the phase of the cavity response is also shown in Fig.~\ref{fig::examples_basicQuantities}. As expected, it features a zero-crossing around the resonance. Another interesting feature is that one can see the boundary of the on-shell region, where $k^2 = k_x^2 + k_y^2$, such that $k_z$ drops to zero. Beyond this boundary $k^2 < k_x^2 + k_y^2$, such that no physical solution for $k_z$ exists. The cavity response is then strictly speaking not defined in this region and all relevant integrals are only over the physical domain, which we refer to as on-shell region.

The incident beam angular spectrum in cavity coordinates is represented by a distorted Gaussian, which has interesting features resulting from the coordinate rotation in wave space. For example, the angular cavity spectrum of the Gaussian pulse is not only asymmetric in the $k_x$ direction, but also the peak maximum does not coincide with the incidence angle, unlike in beam coordinates. This feature is due to the rotation taking place in full wave space, which results in a non-trivially transformation of the angular cavity spectrum (see e.g.~Fig.~\ref{fig::examples_illu_algo_full}). Mathematically, these effects originate in the $F_\delta$-factor in Eq.~\eqref{eq::cav_cavAngSpec_analytic}. Practically, they are important since they influence how optimal settings should be chosen for maximum excitation of the nuclei inside such cavities, as shown in a companion paper \cite{companion_nonlin2024_arxivLetter}. For example, it may be advantageous to tune a collimated beam away from the cavity resonance in order to achieve maximal intensity in the cavity. This unintuitive effect can be directly seen within our analytical solution of the Gaussian beam's angular spectrum in cavity coordinates.

\subsection{Applications}\label{sec::cav_realFocBeam}

\subsubsection{Effect of focusing in optimal cavities for collimated driving}\label{sec::app-opt}
\begin{figure}
	\includegraphics[width=1.0\columnwidth]{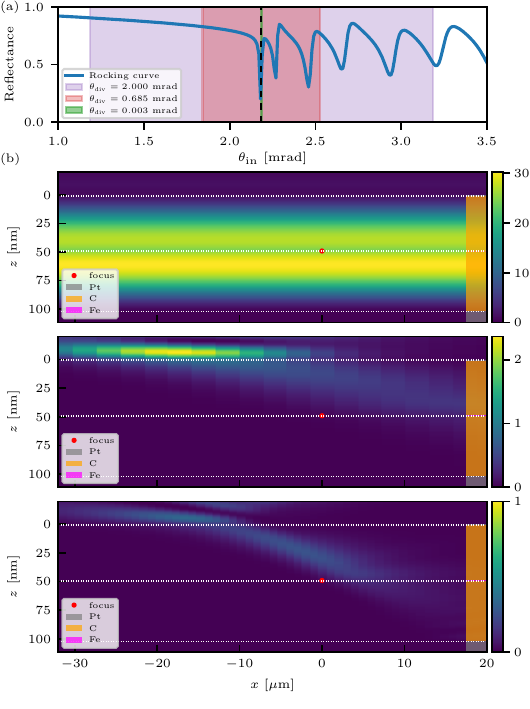}
	\caption{Effect of focusing in a cavity structure that was optimized for maximum field enhancement in the collimated limit \cite{Diekmann2022}. (a) Rocking curve of the cavity indicating the resonance structure. Three different ranges of beam divergences are indicated (see legend). (b) Corresponding resonant field intensities in and around the example cavity for these beam divergences ($\theta_\mathrm{div}=0.1,0.658$ and $2.0\,$mrad; top to bottom). The colored areas and white lines mark the location of the cavity layer planes (see legend). The red dot indicates the focus location of the incident beam. Intensities are normalized to the input pulse peak.}
	\label{fig::examples_realisticBeamDivs_methodShowcase_opt}
\end{figure}
The highly collimated limit of x-ray cavities has been well-studied in the literature \cite{Rohlsberger2005} and in the last decade various experiments have been performed implementing quantum optics phenomena in such a configuration, in particular using the M\"ossbauer resonance of $^{57}$Fe \cite{Rohlsberger2010,Rohlsberger2012,Heeg2013a,Heeg2015a,Heeg2015b,Haber2016a,Haber2017}, but more recently also with electronic resonances~\cite{Haber2019,Huang2021}. Theoretically, optimal cavity structures have been explored using inverse design \cite{Schenk2011PhD,Diekmann2022,Diekmann2022b}. However, it is currently unclear if any of the optimal features survive when focusing the input beam.

In Fig.~\ref{fig::examples_realisticBeamDivs_methodShowcase_opt}, this scenario is considered, that is the effect of focusing in such an optimal cavity is shown. The cavity structure is Pd (1.87\,nm)/C (4.37\,nm)/$^{57}$Fe (1\,nm)/C (3.5\,nm)/Pd with the resonant isotope $^{57}$Fe --- which features a narrow M\"ossbauer transition at 14.4$\,$keV --- and was optimized in \cite{Diekmann2022} towards maximal field enhancement in the resonant layer. The rather large field enhancement of $>30$ is achieved due to a very narrow fundamental mode of the cavity, as can be seen in Fig.~\ref{fig::examples_realisticBeamDivs_methodShowcase_opt}(a). As can be seen in the resulting beam propagation for increasing focusing in Fig.~\ref{fig::examples_realisticBeamDivs_methodShowcase_opt}(b), this also leads to the effect that the beam is quickly and efficiently repelled from the cavity. For high and even for intermediate focusing, almost no intensity enters the cavity and the beam is essentially fully reflect at the cladding with minor evanescent leakage into the cavity. This observation implies that cavities designed for optimal performance at low collimation tend to be particularly bad choices for focused driving. In \cite{companion_nonlin2024_arxivLetter}, we further explore this insight, showing that central design paradigms have to be discarded, but that optimal cavities for focused beams can still lead to good performance. Due to above observation, these cavities turn out to feature rather broad resonances, leading to a higher angular acceptance.

\subsubsection{Focused beam propagation in a multi-layer system}
\begin{figure}
	\includegraphics[width=1.0\columnwidth]{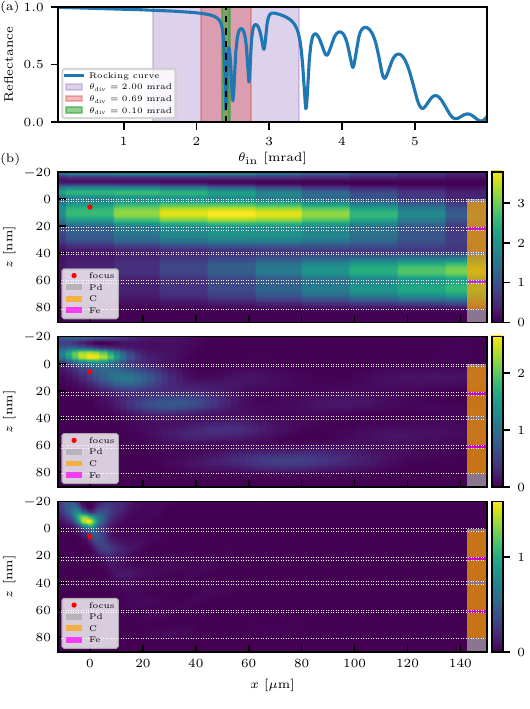}
	\caption{(a) Rocking curve of a multi-layer cavity that was used in \cite{Haber2017} to realize strong coupling between two layers of $^{57}$Fe M\"ossbauer nuclei. Three different ranges of beam divergences are indicated (see legend). (b) Resonant field intensities in and around the example cavity for the beam divergences $\theta_\mathrm{div}=0.1,0.5$ and $2.0\,$mrad (top to bottom). The colored areas and white lines mark the location of the cavity layer planes (see legend). The red dot indicates the beam focus. Intensities are normalized to the input pulse peak.}
	\label{fig::examples_realisticBeamDivs_methodShowcase}
\end{figure}
As a second application, we consider a thin-film x-ray cavity in a stack with many layers to showcase the numerical power of the algorithm. In particular, we investigate the effect of focusing on the beam propagation in the double cavity structure used in the experiment presented in \cite{Haber2017}, where it was shown that such structures enable strong coupling between two nuclear ensembles and result in oscillations between them during the decay phase.

This setup is also interesting from the perspective of nonlinear excitation, since strong coupling is commonly associated with an enhancement of such effects. However, our theoretical argument in Sec.~\ref{sec::excPhase} shows that one has to distinguish between strong inter-ensemble coupling, which affects the decay phase dynamics, and strong coupling to the x-ray field, which affects the excitation phase. The effect reported in \cite{Haber2017} is of the former type. In Fig.~\ref{fig::examples_realisticBeamDivs_methodShowcase}, we investigate if such cavities can nevertheless be used to enhance nuclear excitation when driven with focused x-ray pulses. As for the optimized enhancement cavities in Sec.~\ref{sec::app-opt}, we find that due to the small acceptance angle of these systems supporting multiple modes, the beam is mostly reflected at the outer cladding already for rather low focusing strengths. Interestingly, however, the mode intensity does not decline monotonically between the guiding layers, but can feature a peak after passing one guiding layer with low intensity. Such residual interference effects are useful if optimized for a particular focusing strength, as we utilize in a companion paper \cite{companion_nonlin2024_arxivLetter}.

\section{Conclusion}
In summary, we have developed a semi-analytical algorithm to calculate nonlinear excitation of narrow x-ray transitions by ultra-short pulses in the presence of thin-film cavities. The approach consists of two parts.

First, we reduced the nuclear many-body problem to a simple set of differential equations, which are solved in an approximate analytical fashion by the pulse area theorem. The central assumptions in the approximation hierarchy are that the pulse length is much shorter than the decay time of the x-ray transitions and that the pulse is Fourier limited. The result is that the nuclear excitation after the pulse can be analytically expressed in terms of the resonant Fourier component of the electric field at the transition location.

Second, we developed an algorithm to calculate this Fourier component if the beam propagation is influenced by thin-film x-ray cavities. For the input pulse, we assume a Gaussian beam profile. Since the cavity response can be expressed analytically in parallel wave space, we rotate the angular spectrum of a Gaussian pulse into this coordinate system. The Fourier component of the electric field in the cavity can then be computed by numerically performing one two-dimensional Fourier transform. The algorithm is therefore highly numerically efficient and allows for optimization of the cavity structure, which we exploit in a companion paper \cite{companion_nonlin2024_arxivLetter}.

Finally, we performed various benchmarks and investigated two applications of the propagation of a focused beam in multi-layer cavities, which were optimized for field enhancement \cite{Diekmann2022} or used in strong coupling experiments \cite{Haber2017}.
	
	\begin{acknowledgements}
		The authors would like to thank K.~P.~Heeg, O.~Diekmann, M.~Gerharz, A.~P\'alffy-Bu\ss, R.~R\"ohlsberger and L.~Wolff for valuable discussions and L.~M.~Lohse for comments on an earlier version of the manuscript. DL gratefully acknowledges the Georg
		H. Endress Foundation for financial support and support by the DFG funded Research Training Group ``Dynamics of Controlled Atomic and Molecular Systems'' (RTG 2717).
		
        We note that this paper is based on the doctoral thesis~\cite{LentrodtPhD}. The software and figures in this paper made use of the \textsc{numpy} \cite{numpy2020}, \textsc{scipy} \cite{scipy2020} and \textsc{matplotlib} \cite{matplotlib2007} software packages.
	\end{acknowledgements}

	\newpage
	\appendix
	
	\section{Overview of analytical argument in the paper}\label{app:overview}
\begin{figure*}[p]
		\centering
		\includegraphics[trim={8.4cm 2.4cm 8.7cm 2.6cm},clip,height=0.8\textheight]{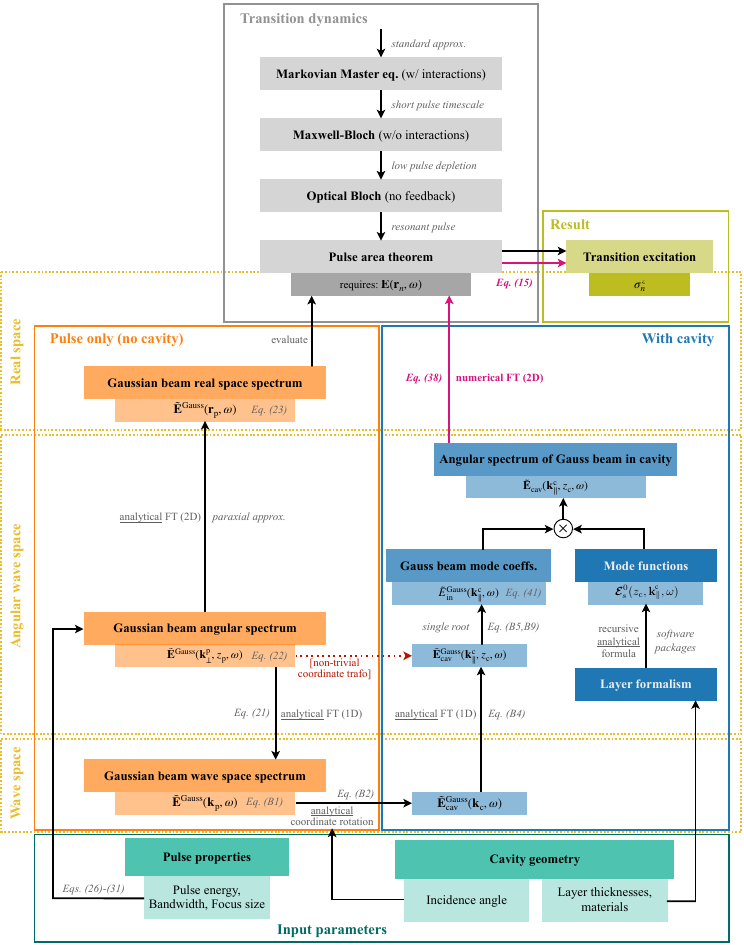}
		\caption{Extended version of Fig.~\ref{fig::examples_illu_algo}, including the approximation hierarchy and analytical derivation of the Gauss beam mode coefficients in cavity coordinates [Eq.~\ref{eq::cav_cavAngSpec_analytic}]. Approximations and equations used in each step are indicated in italics along the arrows. Input parameters: The bottom level (green) constitutes the input parameters, which are subdivided into pulse properties and the cavity geometry. Among the latter, the incidence angle is used to choose the coordinate system for the pulse. Numerics: The numerical part of the algorithm (magenta arrows and connecting boxes) is also shown in Fig.~\ref{fig::examples_illu_algo} and explained in its caption. Derivation: The analytical derivation aims at calculating the Gauss beam mode coefficients necessary for the numerical algorithm, which can be read off from the corresponding angular spectrum in cavity coordinates. The derivation starts with the corresponding angular spectrum in pulse coordinates, which is available analytically as a solution of Maxwell's equations. However, the coordinate transformation between pulse and cavity coordinates is non-trivial (red dotted arrow) in angular wave space, which is needed for the cavity computations (yellow dotted boxes). We therefore first transform to full wave space, where the coordinate transformation is a rotation, and then back to analytically obtain the beam mode coefficients in cavity coordinates for the Gaussian beam as one central result of this paper [Eq.~\ref{eq::cav_cavAngSpec_analytic}]. Approximation hierarchy: The grey boxes show the approximation hierarchy used to analytically solve the nonlinear excitation dynamics of the narrow x-ray transitions during a short driving pulse.}
		\label{fig::examples_illu_algo_full}
	\end{figure*}

	In Fig.~\ref{fig::examples_illu_algo_full}, an extended version of Fig.~\ref{fig::examples_illu_algo} is shown, which also includes the analytical parts of the derivation presented in this paper. The figure is intended to illustrate the logical procedure behind the derivation.
	
	\section{Gaussian beam angular spectrum in cavity coordinates}\label{app::gaussCavCoord}
	
	\subsection{Coordinate rotation in wave space}
	In the following we will express the Gaussian beam field intensity in the cavity coordinate system. For the angular spectrum, a rotation is not possible directly since $x$ and $z$ mix in this transformation. However, the $x$-direction is Fourier transformed in the angular spectrum, such that we have a $k_x$-functionality. Instead, we take the Fourier transform along the $z$-direction, then rotate in full wave space ($k_x$, $k_y$, $k_z$) and Fourier transform back to an angular spectrum afterwards.
	
	The $z_\mathrm{p}$-direction Fourier transform of the angular pulse spectrum Eq.~\eqref{eq::GaussBeam_angSpec} is
	\begin{align}\label{eq::gaussBeam_waveSpec_pulse}
		\tilde{\mathbf{E}}^\mathrm{Gauss}_\mathrm{p}(\mathbf{k}_{\mathrm{p}},\omega) =& \frac{\tilde{\mathbf{A}}(\omega)}{\sqrt{2\pi}}  \frac{w_0^2}{2} e^{-\frac{w_0^2}{4} |\mathbf{k}^\mathrm{p}_\perp|^2} \delta(k^{\mathrm{p}}_z - k^{\mathrm{p},\mathrm{val}}_z) \,,
	\end{align}
	where $k^{\mathrm{p},\mathrm{val}}_z = \sqrt{k^2 - |\mathbf{k}^\mathrm{p}_\perp|^2}$ is the ``on-shell'' wave number in the propagation direction, that is the physical wave number in $z$-direction for a given parallel wave number and frequency.
	
	The wave space spectrum in cavity coordinates is then simply given by
	\begin{align}\label{eq::coordTrafo}
		\tilde{\mathbf{E}}_\mathrm{cav}(\mathbf{k}_\mathrm{c}, \omega) =  \tilde{\mathbf{E}}_\mathrm{p}(\mathbf{k}_\mathrm{p}(\mathbf{k}_\mathrm{c}),\omega) \,,
	\end{align}
	where the pulse wave vector in terms of cavity coordinates is given by a rotation transformation
	\begin{align}\label{eq::coordTrafo_rot}
		\mathbf{k}_\mathrm{p}(\mathbf{k}_\mathrm{c})
		&=
		\begin{pmatrix}
			k^\mathrm{c}_x \sin(\theta_\mathrm{in}) - k^\mathrm{c}_z \cos(\theta_\mathrm{in})
			\\
			k^\mathrm{c}_y
			\\
			k^\mathrm{c}_x \cos(\theta_\mathrm{in}) + k^\mathrm{c}_z \sin(\theta_\mathrm{in})
		\end{pmatrix}\,.
	\end{align}
	Note that in this transformation, the $x$- and $z$-component swap roles in addition to the small $\theta$-rotation. This is due to $z$ in the 
	
	To obtain the angular spectrum in cavity coordinates, we have to perform the inverse Fourier transform along the cavity $z$-direction
	\begin{align}\label{eq::gaussBeam_angFreqFromWave_pulse}
		\tilde{\mathbf{E}}_\mathrm{cav}(\mathbf{k}^\mathrm{c}_\parallel, z_\mathrm{c}, \omega) = \frac{1}{\sqrt{2\pi}}\int dk^\mathrm{c}_z \, e^{ik^\mathrm{c}_z z}\tilde{\mathbf{E}}_\mathrm{cav}(\mathbf{k}_\mathrm{c}, \omega) \,.
	\end{align}
	
	\subsection{Analytical evaluation for the Gaussian beam}
	
	In the following, we calculate the above integral analytically for the case of the Gaussian beam from Sec.~\ref{sec::cav_freeGauss}. The integral is of the form
	\begin{align}\label{eq::gaussBeam_angFreqFromWave_pulse2}
		\tilde{\mathbf{E}}^\mathrm{Gauss}_\mathrm{cav}(\mathbf{k}^\mathrm{p}_\parallel, z_\mathrm{c}, \omega) = \frac{\tilde{\mathbf{A}}(\omega)}{2\pi}  \frac{w_0^2}{2} \int dk^\mathrm{c}_z &e^{ik^\mathrm{c}_z z_\mathrm{c}} e^{-\frac{w_0^2}{4} |\mathbf{k}_\parallel(\mathbf{k}_\mathrm{c})|^2}\nonumber
		\\
		&\times\delta(f_{k_x, k_y}(k^\mathrm{c}_z)) \,,
	\end{align}
	where
	\begin{align}
		f_{k_x, k_y}(k_z) =& \sqrt{k^2 - k_y^2 - [k_x\sin(\theta_\mathrm{in}) - k_z\cos(\theta_\mathrm{in})]^2} \nonumber
		\\
		&- k_x \cos(\theta_\mathrm{in}) - k_z \sin(\theta_\mathrm{in}) \,.
	\end{align}
	This function has two roots at
	\begin{align}
		k^{\pm}_z &= \pm \sqrt{k^2 - k_y^2 - k_x^2} \,.
	\end{align}
	As expected, the roots are therefore on-shell as in the unrotated case. We can then express the $\delta$-function in Eq.~\eqref{eq::gaussBeam_angFreqFromWave_pulse2} as
	\begin{align}
		\delta\left( f_{k_x, k_y}(k_z)  \right) = \sum_{\pm} \frac{\delta\left(k_z - k^\pm_z\right)}{|f'_{k_x, k_y}(k^\pm_z)|} \,,
	\end{align}
	where the prime indicates a derivative along $k_z$. Since our x-ray beam is incident from the top of the cavity and the already small contribution impinging from the bottom due to the angular spread is typically absorbed by sample holders or the side of the cavity in practice, we can perform a single root approximation
	\begin{align}
		\delta\left( f_{k_x, k_y}(k_z)  \right) \approx \frac{\delta\left(k_z - k^+_z\right)}{|f'_{k_x, k_y}(k^+_z)|} \,.
	\end{align}
	The derivative can be evaluated as
	\begin{align}
		f'_{k_x, k_y}(k^+_z) &= \frac{- 1}{\sin(\theta_\mathrm{in}) +\frac{k_x}{\sqrt{k^2 - k_y^2 - k_x^2}}\cos(\theta_\mathrm{in})} \,.
	\end{align}
	Substitution into Eq.~\eqref{eq::gaussBeam_angFreqFromWave_pulse2} then yields the result for the Gaussian beam's cavity angular spectrum Eq.~\eqref{eq::cav_cavAngSpec_analytic}. We note that this form assumes that the focus is at $z_\mathrm{c}=0$. If the focus is moved relative to the cavity, one can either shift the coordinate system accordingly or add an additional phase for the offset, which we employ in our numerical implementation.
	
	\color{black}
    \section{Details on approximations}
    In this appendix, we provide details on critical approximations in the presented approach and further benchmarks to ensure their reliability.
    
    \subsection{The low-depletion approximation in the low-excitation regime}
    Additional insight on the approximations made in the approach in this paper can be obtained by analytic solutions in the low-excitation regime. In particular, the low-depletion approximation is analyzed in the following including a comparison of time and frequency pictures of the dynamics.
    
    \subsubsection{Time picture}\label{sec::timePic}
    In forward scattering geometry, the response function describing the propagation of a field pulse up to depth $z$ through a nuclear medium with resonance frequency $\omega_0$ and effective thickness $b = b(z)$ is given by \cite{HeegPhD}
    \begin{align}
        R(t,z) &= \delta(\tau) - \sqrt{\frac{b(z)}{\tau}}\: J_1(2\sqrt{b(z)\tau})\: e^{-\gamma \tau}\: e^{-i\omega_0 \tau} \Theta(\tau) \nonumber
        \\
        &=  \delta(\tau) + R_d(t,z)\,.
    \end{align}
    Here, $\tau = t-z/c$ includes the retardation of the pulse propagating with velocity $c$, and $R_d(t,z)$ denotes the delayed/scattered contribution. As the initial field, we again consider a short Gaussian x-ray pulse with carrier resonance frequency $\omega_L$
    \begin{align}
        E(t, z=0) = \frac{E_0}{\sqrt{2\pi}\sigma} \: e^{-\frac{t^2}{2\sigma^2} -i\omega_L t} \,.
    \end{align}
    The field propagated to $z$ is then given by
    \begin{widetext}
    \begin{align}
        E(t,z) &= E(t, z=0) \ast R(t,z) \nonumber\\[2ex]
        &= E(\tau, z=0) + \int_{-\infty}^{\infty} E(t-\xi, z=0) \: R_d(\xi, z)\: d\xi \nonumber\\[2ex]
        &= E(\tau, z=0) \nonumber \\
        &\quad- \frac{E_0\, }{\sqrt{2\pi}\sigma} \int_{-\infty}^{\infty} \left[e^{-\frac{(t-\xi)^2}{2\sigma^2} - i \omega_L (t-\xi)}\right] \left[\sqrt{\frac{b(z)}{\xi-\frac{z}{c}}}\: J_1\left(2\sqrt{b(z)\left [\xi-\frac{z}{c}\right]}\right)\: e^{-\gamma (\xi-\frac{z}{c})}\: e^{-i\omega_0 (\xi-\frac{z}{c})} \Theta(\xi-\frac{z}{c})  \right]\: d\xi\,.
    \end{align}
    We perform a variable transformation $\xi \to \xi + z/c$ to obtain
    \begin{align}
        E(t,z) &=  E(\tau, z=0) - \frac{E_0\, }{\sqrt{2\pi}\sigma} \int_{-\infty}^{\infty} \left[e^{-\frac{(t-\xi-\frac{z}{c})^2}{2\sigma^2} - i \omega_L (t-\xi-\frac{z}{c})}\right] \left[\sqrt{\frac{b(z)}{\xi}}\: J_1\left(2\sqrt{b(z)\xi}\right)\: e^{-\gamma \xi}\: e^{-i\omega_0 \xi} \Theta(\xi)  \right]\: d\xi\,.
    \end{align}
    To analyze the peak amplitude of the propagating field, we evaluate it at time $z/c$, on resonance $\omega_L =\omega_0$, considering $\Theta(\xi)$:
    \begin{align}
        E\left(\frac{z}{c},z\right) &=  E(t=0, z=0) - \frac{E_0\, }{\sqrt{2\pi}\sigma} \int_{0}^{\infty} \left[e^{-\frac{\xi^2}{2\sigma^2} }\right] \left[\sqrt{\frac{b(z)}{\xi}}\: J_1\left(2\sqrt{b(z)\xi}\right)\: e^{-\gamma \xi}  \right]\: d\xi \,.
    \end{align}
    \end{widetext}
    We see that the part in the first square bracket constrains the $\xi$-values at which the integrand contributes to a narrow range around $\xi=0$. This range is on the order of the pulse duration, which is very short compared to all relevant time scales of the nuclear dynamics, as described in Sec.~\ref{sec::excPhase}. One can therefore expand the part in the second bracket around $\xi=0$ in leading order to give
    \begin{align}
        E\left(\frac{z}{c},z\right) &\approx  E(t=0, z=0) - \frac{E_0\, }{\sqrt{2\pi}\sigma} \int_{0}^{\infty} \left[e^{-\frac{\xi^2}{2\sigma^2} }\right] \left[b(z) \right] \: d\xi \,.
    \end{align}
    Evaluating the integral, we finally obtain
    \begin{align}
        E\left(\frac{z}{c},z\right) &\approx  E(t=0, z=0) \left[1 - \sqrt{ \frac{\pi}{2} }b(z)\,\sigma \right ]  \,.
    \end{align}
    Now let us consider $\sigma b(z)$ quantitatively. For typical samples, $b\sim \gamma \sim 10^6\,\mathrm{s}^{-1}$. For ultrashort pulses, $\sigma \sim 100\,$fs$\sim 10^{-13}\,$s. Therefore, $b(z) \sigma \sim 10^{-7} \ll 1$ and thus the peak amplitude of the propagated field becomes
    \begin{align}
         E\left(\frac{z}{c},z\right) \approx  E\left(t=0,z=0\right) = \frac{E_0}{\sqrt{2\pi}\sigma} \,.
    \end{align}
    That is the field at depth $z$ {\it at short times} is to a very good approximation the same as the one incident on the target at $t=0$.
    
    \subsubsection{Frequency picture}
    It is interesting to contrast the argument in Sec.~\ref{sec::timePic} with the frequency domain. The response function in the frequency domain is (neglecting a small retardation contribution by setting $\tau\approx t$)
    \begin{align}
        R(\omega,z) = e^{\frac{-i\,b(z)}{\omega - \omega_0 + i\gamma}}\,.
    \end{align}
    The field in the frequency domain as a function of $z$ is obtained by simply multiplying with the response function due to the convolution theorem. On resonance, $\omega = \omega_0$, the attenuation after a propagation distance $z$ therefore yields
    \begin{align}
        E(\omega = \omega_0, z) = E(\omega = \omega_0, z=0)\,e^{-\frac{b(z)}{\gamma}}\,.
    \end{align}
    The attenuation is therefore considerable, since $b(z) \sim \gamma$ for typical sample thicknesses.
    
    \subsubsection{Discussion}
    The solution to this seemingly contradictory result is simple: In the time picture, we in essence could neglect the nuclear response because we only studied the temporal evolution on time scales over which the nuclear dynamics is negligible. Then, the initial pulse essentially remains unchanged. To transfer to the frequency picture, however, the integral over all times is invoked. In the latter case, the nuclear response substantially influences the initial pulse via destructive interference (see also \cite{Kaldun2016,Heeg2017,Heeg2021}), yielding the strong attenuation in the frequency domain. Note that the observation of this narrow resonance indeed requires long observation times due to Fourier relations \cite{Kaldun2016}.
    
    We further note that electronic absorption is caused by processes on fs-as time scales. Therefore, the time separation argument used above is not valid and the initial field is absorbed essentially instantaneously. Electronic absorption is, however, fully included in the algorithm developed in this paper through the cavity response function.

    \subsection{Quantitative benchmark of the low pulse depletion approximation}\label{app:lowPulseDepConfirm}
	In Sec.~\ref{sec::excPhase}, an approximation was used that results in neglecting the feedback effect of photons being absorbed in the nuclear excitation process during the ultra-short pulse onto the pulse profile itself. In this supplementary section, we confirm the validity of this approximation for the optimized cavity targets doped with M\"ossbauer nuclei.
	
	To this end, we calculate the number of photons required for fully inverting the nuclear ensemble according to \cite{companion_nonlin2024_arxivLetter}. Requiring $\Phi^{\mathrm{tot}}_\mathrm{peak}=\pi$ and solving for the photon number gives
	\begin{align}
	N_\mathrm{ph}^{[\mathrm{required}]} = \frac{\sqrt{\pi}\pi^2}{8} \frac{c \varepsilon_0 \hbar w_0^2}{\omega_\mathrm{nuc} \tau_\mathrm{pulse} d^2} \,.
	\end{align}
	This number should then be compared to the number of absorbed excitations after the pulse has passed, which can be estimated as the number of nuclei $N_\mathrm{nuc}$ in the pulse volume. Since we have used the peak inversion above, a conservative estimate --- that is overestimating the number of absorbed photons --- can be obtained via the illuminated volume as
	\begin{align}
	N^{[\mathrm{absorbed}]}_\mathrm{nuc} \approx \pi w_0^2 \frac{t_\mathrm{res} \rho_\mathrm{nuc,res}}{\sin(\theta_\mathrm{in})} \,,
	\end{align}
	where $t_\mathrm{res}$ is the thickness of the resonant layer and $\rho_\mathrm{nuc,res}$ the number density of resonant nuclei. If the Rayleigh length of the pulse is smaller than $\frac{t_\mathrm{res}}{\sin(\theta_\mathrm{in})}$, the number of excited nuclei will be even smaller than in the above estimation.
	
	The low pulse depletion approximation is then valid at full inversion if
	\begin{align}
	\frac{N_\mathrm{ph}^{[\mathrm{required}]}}{N^{[\mathrm{absorbed}]}_\mathrm{nuc}} =  \frac{\sqrt{\pi}\pi}{8} \frac{c \varepsilon_0 \hbar}{\omega_0 \tau_\mathrm{pulse} d^2} \frac{\sin(\theta_\mathrm{in})}{t_\mathrm{res} \rho_\mathrm{nuc,res}} \gg 1 \,.
	\end{align}
	For the parameters considered in Fig.~\figLettInv~of \cite{companion_nonlin2024_arxivLetter}, this quantity can be checked to be $>10^5$ for all considered isotopes. The low pulse depletion approximation is therefore well valid.
    
    \subsection{Full algorithm benchmark via the count rate at synchrotron facilities}
	Due to the current absence of experimental data on nonlinear excitation of nuclei, benchmarking the approximations within our approach beyond the theoretical arguments provided is generally difficult. However, our method also predicts the number of excited nuclei in the low excitation regime. We can then use the experimental experience of scattered photon count rates at synchrotron facilities as an order of magnitude benchmark.
	
	For example, the beamline P01 at PETRA III \cite{Wille2010} offers a photon flux on the sample of $r_\mathrm{ph} \approx 7 \cdot 10^{13} \frac{\mathrm{ph}}{\mathrm{s}}$ and an energy resolution of $b=1\,$meV at a photon energy of $14.4\,$keV (corresponding to the transition energy of $^{57}$Fe). With the pulse separation of $\Delta t_\mathrm{p} = 192\,$ns, this results in
	\begin{align}
		\chi_\mathrm{source} \approx 2.11\cdot 10^{-2} \sqrt{\mathrm{mJ}}
	\end{align}
	according to Eq.~\eqref{eq::chi_source_def}. Together with \cite{companion_nonlin2024_arxivLetter}
	\begin{align}
		\chi_\sigma \approx 8.06\cdot 10^{-12} \frac{\mathrm{m}}{\sqrt{\mathrm{mJ}}} \,,
	\end{align}
	we obtain the pulse area
	\begin{align}
		\Phi \approx \frac{1.7 \cdot 10^{-13}}{w_0}
	\end{align}
	according to Eq.~\eqref{eq::pulse_area_neat}.
	
	At low excitation, which is the case at synchrotron facilities, we can then calculate the number of excited nuclei per shot as $N_\mathrm{exc} = P_\mathrm{exc} N_\mathrm{illu}$, where $P_\mathrm{exc}$ is the probability of excitation and $N_\mathrm{illu}$ is the number of illuminated nuclei. In the low excitation regime, from which no deviations have been found so far at synchrotron facilities, we can write $P_\mathrm{exc} = 1 - \cos(\Phi)\approx \frac{\Phi^2}{2}$. Assuming a forward scattering geometry with a typical target of $t_\mathrm{target}=$1$\,\mu$m single line iron enriched in $^{57}$Fe, we further find $N_\mathrm{illu} \approx t_\mathrm{target} \pi w_0^2 \rho_\mathrm{N}$, where $\rho_\mathrm{N}\approx 4.2 \cdot 10^{28}\,\mathrm{m}^{-3}$ is the number density of the nuclei and we have roughly estimated the illuminated area as $\pi w_0^2$. Overall, this results in $N_\mathrm{exc} \approx 2 \cdot 10^{-3}$. Together with the excitation lost to internal conversion and assuming a good detection efficiency, this results in an estimation for the number of detected photons per shot of $N_\mathrm{det} \approx 2 \cdot 10^{-4}$ or one detected photon every $\sim$5000th pulse. This rough estimate matches the experimental experience at such facilities \cite{Wille2010} rather well, confirming that our approximations are at least roughly correct, excluding significant offsets which could potentially arise e.g.~from an incorrect treatment of collectivity. Together with the theoretical arguments provided, this benchmark provides a solid foundation for using our approach as a predictive theory.

    \color{black}
	
	\bibliographystyle{myprsty}
	\bibliography{library}
	
\end{document}